\documentclass[%
 preprint,
nofootinbib,
 amsmath,amssymb,
 aps,
prb,
showkeys
]{revtex4-1}

\usepackage{amssymb}
\usepackage{graphicx}
\usepackage{dcolumn}
\usepackage{bm}

\usepackage{textcomp}

\newcommand{\rev}[1]{#1}

\begin{document}




\title{Lifshitz theory of wetting films at three phase coexistence: \\ The case
of ice nucleation on Silver Iodide (AgI)}%


\author{Juan Luengo-M\'arquez}
\affiliation{%
Departamento de Qu\'imica F\'isica, Universidad Complutense de Madrid
}%
\author{Luis G. MacDowell}%
\affiliation{%
Departamento de Qu\'imica F\'isica, Universidad Complutense de Madrid
}%




\date{\today}

\begin{abstract}  
\paragraph{Hypothesis:} As a fluid approaches three phase coexistence, adsorption may take
place by the successive formation of two intervening wetting films. The equilibrium thickness of these wetting layers is the result of a delicate balance
of intermolecular forces, as dictated by an underlying \rev{interface potential}. 
The van der Waals forces for the two variable adsorption layers may be 
formulated exactly from  Dzyaloshinskii-Lifshitz-Pitaevskii theory, and
analytical approximations may be derived that extent well beyond the validity
of conventional Hamaker theory.

\paragraph{Calculations:} We consider the adsorption equilibrium of water vapor
 on Silver Iodide where both ice and a water layers  can form
simultaneously and compete for the vapor as the triple point
is approached.  We perform numerical calculations of Lifshitz theory
for this complex system and work out analytical approximations which
provide quantitative agreement with the numerical results.
 
\paragraph{Findings} 
At the  three phase contact line between AgI/water/air,
surface forces promote growth of ice both on the AgI/air and the
water/vapor interfaces, lending support to a contact nucleation mode
of AgI in the atmosphere.  Our approach provides a
framework for the description of adsorption at three phase coexistence, and
allows for the study of ice nucleation efficiency on atmospheric
aerosols.
\end{abstract}


\maketitle





\section{Introduction}

The preferential adsorption of gases and liquids on an inert solid
substrate is ubiquitous in colloidal science, and is often accompanied by the 
formation of thick wetting films that span from a few nanometers
to several microns as two phase coexistence is approached 
\cite{derjaguin87,degennes04,israelachvili11,schick90}.
The behavior of the equilibrium layer formed can be fully
characterized by an interface potential, $g(h)$, which measures
the free energy as a function of the film thickness, $h$
\cite{dietrich88,schick90,degennes04,henderson05}. For
thick films, $g(h)$ is often dominated by long range van der Waals forces
that can be described rigorously with the celebrated
Dzyaloshinskii-Lifshitz-Pitaevskii (DLP)  theory 
\cite{dzyaloshinskii61}.  

A more complex situation arises when the adsorbed wetting film can
further segregate and form a new layer between the solid substrate
and the mother phase as three phase coexistence is approached
\cite{takeya06,aspenes10,nakane19,bostrom19,antonov14,rafai07,hartlein08}.
This can be quite generally the case for substrates in contact
with multicomponent mixtures.  Examples include the industrially relevant 
formation
of clathrate hydrates from aqueous solutions of oil or carbon dioxide 
\cite{takeya06,aspenes10,nakane19,bostrom19},
biologically relevant aqueous solutions of Dextran and
Bovine Serum Albumin \cite{antonov14}; or theoretically relevant model solutions
such as ethanol/n-alkane or lutidine/water mixtures \cite{rafai07,hartlein08}.
The wetting problem now becomes considerably more complex, as the
system can exhibit two thick layers of size, say, $l$ and $d$, respectively,
that are bounded by the substrate and the mother phase and can feed one
from the other depending on the prevailing thermodynamic conditions.  

In practice, this complex scenario can be realized for a very
relevant one component test system, namely, atmospheric supercooled water vapor 
in close proximity to the triple point \cite{dash06}. As ice
nucleates on the surface of inorganic aerosols, the resulting
ice/water vapor interface is exposed. This surface is actually
 a complex system exhibiting a thin premelted
water layer, usually referred as \textit{Quasi-Liquid Layer} (QLL). Its
properties largely condition several phenomena related to ice, such as the electrification of storm
clouds, frost and snowflakes formation, or ice  
skating \citep{rosenberg05,dash06,slater19,nagata19}. 
Hereby the surface Van der Waals forces
play a crucial role in the stabilization of a thick
QLL\citep{elbaum91b,fiedler20}. 

\begin{figure}[h]
\includegraphics[scale=0.75]{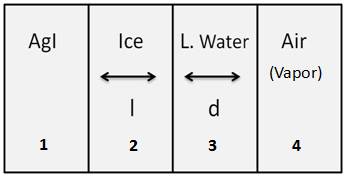}
\caption{Geometry of  layered planar systems studied in this work.
An inert substrate, AgI, in contact with water vapor (air), close
to three phase coexistence can exhibit intervening phases of ice,
with thickness $l$ and water, with thickness $d$. Since the system
is subject to material equilibrium each phase feeds from the other. The
equilibrium layer thickness of $l$ and $d$ are dictated by
the bulk free energies and the \rev{interface potential}.
For the sake of generality and conciseness of notation, we denote
the phases AgI, ice, water, air as '1', '2', '3', '4', respectively.  }
\label{figmedia}
\end{figure}

On the other hand, silver iodide may be used as a nucleus for ice
formation\citep{vonnegut47}, with applications in cloud seeding to
induce rainfall over wide areas. The conventional belief was that the capability
of the AgI to influence the growth of ice was connected to their lattice match.
Nonetheless, recent studies\citep{marcolli16} have found  substances with
similar structures that do not promote nucleation. Thus they relate the faculty
of the AgI to serve as ice nucleating agent to the charge distribution in the
substrate \cite{fraux14,zielke15,glatz16}. The present work aims at elucidating the role
of Van der Waals forces in the ice nucleating activity of silver iodide, by
describing precisely the Van der Waals interactions in the framework of the
Lifshitz theory, applied to the AgI - Ice - Liquid Water - Air system (Fig. \ref{figmedia}).

The conventional view is that the van der Waals interactions
between two surfaces 
result from the summation of  forces between pairs of particles. The interaction
coefficients are then added into the Hamaker coefficient, $A_{Ham}$, and the
interaction for a layered planar system provides an energy $A_{Ham}/(12\pi
h^{2})$, being $h$ the distance between the interacting media
\cite{israelachvili11}. The Lifshitz theory of the Van der Waals forces goes
beyond the pairwise summation, by treating the Van der Waals forces through
continuum thermodynamic properties of the polarizable interacting
media\citep{parsegian05}. As a particularly significant result of
the detailed electromagnetic treatment, there appears a range of
distances where the
leading order interactions become {\em retarded} and decay as $1/h^{3}$.
Accordingly, the effective interaction coefficient itself, $A_{Ham}$, is no
longer a constant, but becomes a function of the film thickness as well.
Then the generalized expression for the Van der Waals free energy 
becomes\citep{parsegian05}:
\begin{equation}
g_{VdW}(h) = -\frac{A_{Ham}(h)}{12\pi h^{2}}
\label{hamfunc}
\end{equation}
In the study of ice growing at the AgI/air interface, the simple description
based on interactions between AgI and air across ice is not valid. Because of
the existence of premelting, we also expect the formation of a fourth layer of
water in between ice and air. The extension of Lifshitz theory to four media of
which one is variable has been known for a long time
\citep{ninham70c,podgornik03,parsegian05}.
However, in the atmosphere the water layer can grow at the expense of the ice
layer, depending on the prevailing conditions. Therefore both the thickness of
the ice layer (l) and that of the premelted layer (d) must be taken as
variables \cite{esteso20}.

In the next section we first formulate the problem of three phase adsorption
equilibrium generally, and discuss the conditions for wetting. In Section  \ref{lifshitz} we
present the theoretical framework of van der Waals forces in its modern 
formulation  \cite{vankampen68,ninham70b},  generalize the results to a system
of two layers of variable thickness sandwiched between infinite bodies and describe new accurate 
analytical approximations for the van der Waals forces. The parameterization of dielectric properties for AgI, water and ice
 that is required as input into DLP theory is described in Section \ref{sec:model}. In Section \ref{sec:results} we present our results. 
These are structured as follows: first, we test the numerical approximations performed to achieve
closed analytical expressions. Secondly, the theory is exploited to describe the van
der Waals forces of the complex system AgI/Ice/water/air, and their role in
the ice nucleating efficiency of AgI.  The conclusions of our work are presented in 
Section \ref{sec:conclusion}.

\section{Adsorption equilibrium at three phase coexistence}

\label{adsorb}

Consider a fluid phase, say, medium '4', in contact with an inert substrate,
say, medium '1'. Furthermore, consider that  phase, '4', extends
up to a finite but very large distance $L\to\infty$. Accordingly, '4' serves
as a heat and mass reservoir that fixes the temperature and chemical potential
of the full system (of course, if '4' is a multicomponent mixture, it fixes
the chemical potential of each of its components). Quite generally, the density 
of phase '4' in the vicinity
of the substrate is not that found in the bulk phase well away from the wall.
Particularly,
consider that phase '4' is approaching three phase coexistence, such that
two additional bulk phases '2' and '3' are slightly metastable. In a bulk
system in the thermodynamic limit, slightly metastable means that these phases
are not observed at all. However, close to an inert substrate, surface forces
can change this situation, as one of the metastable phases could preferentially
adsorb between the substrate and the mother phase '4'. By the same token,
once, say, phase '2' has adsorbed preferentially in between the substrate
and phase '4', the third phase '3', could adsorb preferentially in between
'2' and '4', leading to a two layer system of phases '2' and '3' in between
the substrate '1' and the mother phase '4'.

The question is then what sets the equilibrium film thickness of the intervening
layers, '2' and '3', with thickness $l$ and $d$, respectively.

Since we assume the system is at fixed temperature and chemical potential
as dictated by the semi-infinite phase '4', the equilibrium states will be such
that the surface grand free energy, $\omega$ is a minimum \cite{callen85}. In the mood
of capillarity theory, we assume the total free energy is that of infinitely
large bulk systems, plus the cost to form each of the interfaces. Following
Derjaguin, however, we also need to account for the effective interaction
between interfaces separated by a finite distance, via a generalized
\rev{interface potential} $g_{1234}(l,d)$ \cite{derjaguin87}. With this in mind, 
we find:
\begin{equation}
   \Delta \omega(l,d) = -(p_2-p_4) l - (p_3-p_4) d 
    + \gamma_{12} + \gamma_{23} + \gamma_{34} + g_{1234}(l,d)
    \label{totfree}
\end{equation} 
where $\Delta\omega=\omega+p_4L$  is an excess over the bulk free energy
of a system filled with phase '4' only. 
Here, $p_i$ is the bulk pressure of phase $i$ at the fixed temperature 
and chemical potential of phase '4', and $\gamma_{ij}$ is the surface tensions between phase
$i$ and phase $j$. 
The first two terms in Eq.\ref{totfree} account for the bulk free energy of
the system; the next three correspond to the free energy to form the interfaces
separating infinitely large bulk phases; and finally, $g_{1234}(l,d)$ accounts
for the missing interactions due to the finite extent of phases '2' and '3'.
By this token it follows immediately that the \rev{interface potential} is defined
such that in the limit $l\to\infty$ and $d\to\infty$, $g_{1234} = 0$.

This result differs from  previous work by additive constants \cite{esteso20}, 
which merely correspond to a different choice of reference state.
This does not change the equilibrium condition for $l$ and $d$, which are obtained by equating
to zero the partial derivatives of the free energy with respect to $l$ and $d$
\cite{esteso20}. However, as written here Eq.\ref{totfree} allows  to rationalize 
in a nutshell the adsorption equilibrium of a three phase system. 

In practice, we will be concerned with the situation where the system
is exactly at bulk three phase coexistence, such that $p_2=p_3=p_4$. The
above equation then allows us to generalize the condition for wetting
at three phase coexistence.

If Eq.\ref{totfree} has an extremal point at \rev{finite} $l=l_0$ and $d\to\infty$,
\rev{the system has three bulk phases, namely, phases '1', '3' and '4',
with a microscopically thin adsorption layer of phase '2' intervening
between '1' and '3'.  Accordingly, it exhibits only two interfaces.
One is a surface
enriched interface between phases '1' and '3', with a surface tension
$\gamma_{13}$ and the other is the interface separating bulk phases '3' and '4',
with surface tension $\gamma_{34}$.  This has an
overall free energy cost $\Delta \omega(l,d) = \gamma_{13}+\gamma_{34}$. 
Equating this result to Eq.\ref{totfree}, we find}:
\begin{equation}
   \lim_{d\to\infty} g(l_0,d) = \gamma_{13} - \gamma_{12} - \gamma_{23}
\end{equation} 
\rev{When $g(l_0,d\to\infty)$ vanishes, this equation}
provides the wetting condition for phase \rev{'2'} intervening between
the substrate and  phase '3'. \rev{If the condition is obeyed for 
finite $l_0$, one obtains a first order wetting transition whereby 
the thickness of the adsorption layer
of phase '2' jumps discontinually from a finite value $l_0$, below the 
wetting transition,
to an infinite value  above the wetting transition.
Alternatively, if the condition is met only as $l_0\to \infty$, it
corresponds to a second order wetting transition
\cite{dietrich88,schick90,degennes04}.}

Likewise, if Eq.\ref{totfree} has extrema at $l\to\infty$ and finite $d=d_0$,
the system forms interfaces with a cost $\gamma_{12}$ and
$\gamma_{24}$, and we find:
\begin{equation}
   \lim_{l\to \infty} g(l,d_0) = \gamma_{24} - \gamma_{23} - \gamma_{34}
\end{equation} 
\rev{In the case that $g(l,d_0)=0$, this}
provides the wetting condition for phase '3' intervening between
phase '2' and '4'. 

Finally, if Eq.\ref{totfree} has an extrema at finite and small  $l=l_0$ and
$d=d_0$ (including the case where both $l_0=d_0=0$), we find:
\begin{equation}
   g(l_0,d_0) = \gamma_{14} - \gamma_{12} -
 \gamma_{23} - \gamma_{34}
\label{wettot}
\end{equation}
\rev{When  $g(l_0,d_0)$ vanishes, it becomes equal to the interface potential of
a system with infinitely thick layers of phase '2' and '3' intervening
between phase '1' and phase '4', which is null by construction. Accordingly, 
$g(l_0,d_0)=0$ is the condition for a first order wetting transition
where the thickness of '2' and '3' jumps discontinually
from finite thicknesses, $l_0$ and $d_0$, to infinite values.
}

The above results serve to illustrate the crucial significance of
the \rev{interface potential} $g(l,d)$ \rev{as a means to characterize wetting
behavior}. Not only it dictates the allowed equilibrium
values of layer thickness \cite{esteso20}, but it also embodies all allowed wetting
conditions in the system. 

The \rev{interface potential} consists of contributions
of different nature \rev{\cite{israelachvili11,starov09} }. First,
a structural contributions , which is short range, as it decays in the
length-scale of the bulk correlation length of a few molecular diameters, 
conveys information on the packing correlations between the hard core of the molecules 
\rev{\cite{tarazona85,chernov88,henderson94,henderson11,hughes15,hughes17}}; 
Second, van der Waals contributions, which are long ranged and result from
spontaneous electromagnetic fluctuations of the media
\rev{\cite{dzyaloshinskii61,parsegian05}}.  Additionally, charged systems will also
have electrostatic contributions, with a decay that is given by the
Debye screening length \rev{\cite{israelachvili11,starov09}}. 
\rev{Because of this complicated superposition of interactions at different
length scales, predicting the full interface potential is extremely challenging.
However, in the absence of electrostatic interactions, it has been shown 
that at length scales of several nanometers, van der Waals contributions
dominate completely over the short range structural forces
\cite{sabisky73,blake75,israelachvili78}. Accordingly,
the behavior of thick wetting films may be determined from the
evaluation of van der Waals forces alone \cite{fenzl03}.

In the next section we discuss the calculation of this
dominant contribution as predicted from the highly accurate DLP theory
\cite{dzyaloshinskii61,parsegian05}.}

\section{Lifshitz theory of Van der Waals forces}

\label{lifshitz}

The starting point of this section is the general result for the surface free energy
embodied between two semi-infinite planar bodies, $L$ and $R$, separated by an arbitrary
number of layers (\rev{$m_1,m_2,\cdots$}) due to van der Waals forces:\cite{ninham70b} 
\begin{equation}
   g_{Lm\dots R} = \frac{k_{B}T}{2 \pi}{\sum_{n=0}^{\infty}}^\prime\int_{0}^{\infty}\rho
\hspace{0.1cm} d\rho \ln(D^{E}_{R\rev{m_1}\dots L}D^{M}_{L\rev{m_1}\dots R})
\label{general_lifshitz}
\end{equation}
\rev{where the prime next to the sum indicates that the term $n=0$ has
an extra factor of $1/2$; while
$D^{E}(n,\rho)$ and $D^M(n,\rho)$ are functions that equated to zero provide
the dispersion relations of the standing waves of
electric and magnetic modes in the system}.  The integral is performed over transverse components of the momentum, 
and the sum is performed over an infinite set of discrete Matsubara frequencies
\rev{$\xi_{n} = \frac{2\pi k_{B}T}{\hbar}n$,
with c, the velocity of light,  $\hbar$, Planck's constant in units of angular
frequency and  
} $k_B$, Boltzmann's constant. We  
assume a temperature $T=273.15$~K set to the triple point of water.

The dispersion relations \rev{implied by $D^{E}(n,\rho)$ and $D^M(n,\rho)$} 
depend on the 
specific geometry of the system. For a
layered planar system composed of two bulk bodies  separated
by a dielectric with thickness '$h$', the result is well known \cite{ninham70b}.
In this work we deal instead with bulk bodies (say, $L=1$ and $R=4$)
separated by two different layers (say \rev{$m_1=2$, $m_2=3$}) with variable thicknesses
'$l$' and '$d$', respectively (Fig.\ref{figmedia}).
The corresponding \rev{$D^{E,M}(n,\rho)$} of the four media system was anticipated by
Esteso et al.\citep{esteso20}, and is derived in the
{\em Supplementary Material 1}. The result is:
\begin{equation}
\begin{array}{ccc}
   D^{E,M}_{1234} & = & 1 - \Delta_{12}^{E,M}\Delta_{32}^{E,M}e^{-2 \rho_{2}l} -
\Delta_{23}^{E,M}\Delta_{43}^{E,M}e^{-2 \rho_{3}d} \\
    & & \\
           & &
  - \Delta_{12}^{E,M}\Delta_{43}^{E,M}e^{-2 \rho_{2}l-2 \rho_{3}d} 
 \end{array}
\label{disp_rel1234}
\end{equation}
Assuming from now on that all magnetic susceptibilities are equal to one\citep{parsegian05}, 
the $\Delta_{ij}(n,\rho)$ functions have the form
\begin{equation}
   \Delta_{ij}^{E}(n,\rho) = \frac{\rho_{j} - \rho_{i}}{\rho_{j} + \rho_{i}}
   \hspace{0.5cm} \Delta_{ij}^{M}(n,\rho) = \frac{\epsilon_{i}\rho_{j} - \epsilon_{j}\rho_{i}}{\epsilon_{i}\rho_{j} + \epsilon_{j}\rho_{i}}
\label{deltas}
\end{equation}
where $\rho_{i}^{2} = \rho^{2} + \frac{\epsilon_{i}\xi_{n}^2}{c^{2}}$, and
\rev{$\epsilon_{i}=\epsilon_i(i\xi_n)$}, the \rev{frequency dependent} dielectric function 
\rev{or complex permitivity} of the medium $i={1,2,3,4}$ is evaluated
at the complex Matsubara frequency $i\xi_{n}$ \rev{(c.f. Eq.\ref{damped_dielec} and
Ref.\cite{parsegian05} for further details on the definition of
the dielectric function)}.

The related result for the interaction between a plate coated with
a layer of {\em fixed} thickness $l$ and another plate at a variable distance $d$ has
been known for a long time (c.f. \cite{ninham70c,podgornik03,parsegian05}). Here, Eq.\ref{general_lifshitz} together
with Eq.\ref{disp_rel1234} generalize this result for the case were {\em both}
$l$ and $d$ are variable. 
\rev{Whereas there is no conceptual difference between the two expressions,
in practice calculations for a single variable thickness allow one to circumvent
the cumbersome derivation that is required to calculate explicitly the
dispersion relation for two media of variable thickness. 
In the Supplementary Material 1, we show that the difference between the 
two expressions amounts merely to a normalization constant setting the zero
of energies. Unfortunately, the simplification is at the cost of loss
of generality of the resulting expression.   In practice, we
find that the more general result  Eq.\ref{disp_rel1234} 
may be expressed in  tractable form
after some lengthly manipulations (c.f. Supplementary Material 1).
In this way},
the layer thickness $l$ and $d$ stand now on the same footing.
 The \rev{general} result also satisfies a
number of desirable physical properties, and is consistent with expectations
for the limiting cases where the layers either become infinitely thick or
vanish altogether. 

Firstly, one notices that in the limit where both $d$ and $l$ tend
to infinity, $D^{E,M}_{1234}\to 1$, so that the \rev{interface potential}
vanishes, as implied in the discussion of the previous section.

Secondly, in the limiting case where either
$d\to\infty$ or $l\to\infty$, one readily finds from Eq.\ref{disp_rel1234} 
that we recover the known dispersion relation for a single layer separating
two plates, as expected. Particularly, for $d\to\infty$,
$D^{E,M}_{1234}\to D^{E,M}_{123}$. As a result, it follows:
\begin{equation}
   \lim_{d\to\infty} g_{1234}(l,d) = g_{123}(l)
\label{dinf}
\end{equation} 
which corresponds to the \rev{interface potential} for layer '2' \rev{separating}
semi-infinite bodies '1' and '3'.  Likewise, for $l\to\infty$,
we recover the \rev{interface potential} for layer '3' \rev{separating} semi-infinite
bodies '2' and '4':
\begin{equation}
   \lim_{l\to\infty} g_{1234}(l,d) = g_{234}(d)
\label{linf}
\end{equation} 
With some additional algebraic work (c.f. Supplementary Material 1), we also find for the opposite limit
of vanishing thickness that:
\begin{equation}
   \lim_{d\to 0} \left ( g_{1234}(l,d) - g_{234}(d) \right ) = g_{124}(l)
\label{dzero}
\end{equation} 
\begin{equation}
   \lim_{l\to 0} \left ( g_{1234}(l,d) - g_{123}(l) \right ) = g_{134}(d)
\label{lzero}
\end{equation} 
These four equations illustrate the great generality embodied in the
dispersion relation Eq.\ref{disp_rel1234} and serve as a check of consistency for the numerical
calculation of $g_{1234}(l,d)$.

\subsection{Analytic approximation for surface Van der Waals forces}

\label{sec:theory}

As evidenced by the above results, $g_{1234}(l,d)$ contains information
on $g_{123}(l)$ and $ g_{234}(d)$. This can be shown explicitly
upon linearization of the logarithmic term in Eq.\ref{general_lifshitz}. The
resulting expression can be readily interpreted as given by:
\begin{equation}
g_{1234}(l,d) = g_{123}(l) + g_{234}(d) + \Delta g_{1234}(l,d)
\label{decomposition}
\end{equation}
Accordingly,  $g_{1234}(l,d)$  may be given as a sum of $g_{123}(l)$ 
and $g_{234}(d)$, plus a correction $\Delta g_{1234}(l,d)$ which accounts for
the indirect interaction of the two macroscopic bodies across the layers.

Despite this simplification, the expressions for the three media potential
still remain very difficult to interpret intuitively. In the next section
we derive analytical formulae which allow to interpret $g_{1234}(l,d)$ 
easily and serve also as an accurate means to calculate the free energy
efficiently.

The results are derived along the same lines  as a theory for the calculation of 
interface potentials in three media reported recently \citep{macdowell19}, 
so we here briefly sketch the solutions and discuss the mathematical
details of the derivation in the Supplementary Materials section 2-7.

\subsubsection{Analytical approximations for the three media contributions}

\rev{Eq.\ref{decomposition} is obtained after linearization of the
four media interface potential. The first two terms in that
expression correspond exactly to linearized three body interface potentials,
of the form:}
\begin{equation}
   g_{123}(h) = -\frac{k_{B}T}{8\pi h^{2}}{\sum_{n = 0}^{\infty}}^\prime
    \int_{r_{n}}^{\infty} dx\, R(n,x) x e^{-x}
\label{threemedsimp}
\end{equation}
where the standard change of variables  $\rho_m\to\rho$ and $x\to 2h\rho_m$
has been performed \cite{parsegian05} (\textit{Supplementary Material 2}),
$R(n,x) = \Delta_{12}^{E}\Delta_{32}^{E} +
\Delta_{12}^{M}\Delta_{32}^{M}$, and
the lower limit of the integral is $r_{n} = 2h\sqrt{\epsilon_{2}}\xi_ {n}/c$.
\rev{Because of the change of variables,  the functions
$\Delta_{ij}^{E,M}(n,x)$ are evaluated according to Eq.\ref{deltas}, with
$\rho_i^2$ replaced by 
$x_i^2= x^{2} + (\epsilon_{i}-\epsilon_{2})(2h\xi_{n}/c)^{2}$}

Although this expression is rather cumbersome and has been traditionally solved numerically, 
it has been shown recently that very accurate analytical approximations may be 
obtained using problem adapted  one-point
Gaussian quadrature rules \citep{macdowell19}. The one-point Gaussian
quadrature allows the
transformation of an integral $\int f(x) w(x) dx$ without known primitive, 
into the product of an
elementary integral $\int w(x) dx$ and the function $f(x)$
evaluated at the quadrature point, $x_{1}$. Essentially, this corresponds
to the application of the mean value theorem, with the one-point Gaussian quadrature
rule exploited as a means to estimate $x_1$ (detailed in \textit{Supplementary
Material 3}).

\paragraph{First Gaussian Quadrature Approximation}

In order to solve the non-trivial integral of Eq.\ref{threemedsimp},
we identify the function $R(n,x)$ to $f(x)$, and
set $w(x) = x \hspace{0.05cm}e^{-x}$ as the weight function.
Applying a one point Gaussian quadrature (\textit{Supplementary Material 4}),
we find the three media \rev{interface potential} splits into
$g_{123}(h) = g_{123}^{\xi_{n} = 0}(h) + g_{123}^{\xi_{n} > 0}(h)$.
The first term, corresponding to $n=0$, is the well known
expression:
\begin{equation}
g_{123}^{\xi_{n} = 0}(h) = -\frac{k_{B}T}{16\pi h^{2}}\left(\frac{\epsilon_{1}-\epsilon_{2}}{\epsilon_{1}+\epsilon_{2}}\right)\left(\frac{\epsilon_{3}-\epsilon_{2}}{\epsilon_{3}+\epsilon_{2}}\right)
\label{zero_term_main}
\end{equation}
The finite frequency contribution $g_{123}^{\xi_{n} > 0}(h)$ is:
\begin{equation}
g_{123}^{\xi_{n} > 0}(h) = -\frac{k_{B}T}{8\pi h^{2}}\sum_{n=1}^{\infty}R(n,x_{1})(1 + r_{n})e^{-r_{n}}
\label{rest_term_main}
\end{equation}
where $R(n,x_1)$ is evaluated at the quadrature point $x_{1} = (2 + 2r_{n} +
r_{n}^{2})/(1 + r_{n})$.
We denote the result of Eq.\ref{zero_term_main} and Eq.\ref{rest_term_main} as the
First Gaussian Quadrature Approximation (FGQA).

\paragraph{Second Gaussian Quadrature Approximation}

In the FGQA, the finite frequency term remains as an awkward infinite
series. However, at ambient temperature the terms in the series
remain sufficiently close to each other that the sum can be \rev{approximated to} an 
integral. Indeed, using the Euler-MacLaurin formula, we find that
corrections to the integral approximation are negligible 
(\textit{Supplementary Material 5}). Accordingly, we introduce the
integration variable  $\nu_{n} = \nu = \nu_{T}n$, with $\nu_{T} =
2\sqrt{\epsilon_2}\xi_{T}/c$ and $\xi_{T}=2\pi k_{B}T/\hbar$,
\rev{approximate the sum to an integral}
and apply again a one point Gaussian quadrature rule (c.f.
   Ref.\citep{macdowell19} and \textit{Supplementary Material 6}). 
The outcome  is the quadrature point $\nu^{*}=\nu_{T} + \nu_{\infty}\xi$, 
where $\xi$ is an adimensional factor
\begin{equation}
\xi = \frac{(\nu_{T}h + 1)(\nu_{\infty}h+1) + 2\nu_{\infty}h}{(\nu_{\infty}h+1)^{2}(\nu_{T}h + 1)+(\nu_{\infty}h + 1)\nu_{\infty}h}
\label{adimensional_factor_main}
\end{equation}
together with the approximate expression for the free energy as:
\begin{equation}
   g_{123}^{\xi_{n} > 0}(h) = -\frac{c\,\hbar \nu_{\infty} }{32\pi^{2} h^{2}}\widetilde{R}^{*}_{\xi}\widetilde{F}
\label{third_changed_integral_main}
\end{equation}
where 
$\widetilde{R}^{*}_{\xi} = \epsilon_{2}^{-1/2}j_{2}^{-1}
\widetilde{R}(\nu^{*},x_{1})e^{\xi}$,
$j_{2} = \left( 1 + \frac{1}{2}\frac{d\ln \epsilon_{2}}{d \ln \xi}\right)$
and
\begin{equation}
\widetilde{F} = \frac{(\nu_{T}h + 1)(\nu_{\infty}h+1) + \nu_{\infty}h}{(\nu_{\infty}h+1)^{2}} e^{-\nu_{T}h}
\label{wide_F}
\end{equation}
Here, $\nu_{\infty}$ is a parameter which is introduced to ensure convergence
of the one point quadrature rule. Physically, it  corresponds to a wave-number
in the range at which the dielectric functions converges to the permitivity of the vacuum. For
practical purposes, $\nu_{\infty}$ is obtained so that  $g_{123}^{\xi_{n} > 0}(h)$ 
matches the exact Hamaker constant of the system \cite{macdowell19}. \rev{In practice, the factor $j_{2}$ may be approximated to 1 in the calculations.}

Although the expression for $g_{123}^{\xi_{n} > 0}(h)$ appears rather lengthy,
$\widetilde{R}^{*}_{\xi}$ depends on $h$ very weakly.
Accordingly, the function $\widetilde{F}$ conveys the leading order correction
of the free energy with respect to the simple Hamaker power law. 
Inspection of Eq.\ref{wide_F} shows that
$g_{123}^{\xi_{n}>0}$ depends on the two inverse length scales, 
$\nu_{\infty}$ and $\nu_T$.  When
$h\ll\nu_{\infty}^{-1}$, $\widetilde{F}$ becomes a constant, and
$g_{123}^{\xi_{n}>0}\propto h^{-2}$.  This is the Hamaker limit of 
non-retarded interactions. For $\nu_{\infty}^{-1}\ll h\ll\nu_{T}^{-1}$, 
$\widetilde{F}$ falls as $~1/h$, so that
$g_{123}^{\xi_{n}>0}\propto h^{-3}$, which corresponds to the Casimir regime
of retarded interactions. Finally, when $h\gg\nu_{T}^{-1}$, the evolution of 
the interaction is dominated by the exponential $e^{-\nu_{T}h}$
\rev{which vanishes altogether for large $h$. This is a finite temperature
effect, not included in the original formulation of Casimir for the interaction
between perfect conductors. Indeed, for $T \to 0$, $\nu_T\to 0$, so that
the Casimir retarded regime persists up to infinitely large distances, as
expected.

In practice, the suppression
of retarded interactions at ambient temperature is not of great practical relevance, because
$\nu_T^{-1}$ is of the order of the micrometer at ambient temperature, a
distance where the surface interactions is at the limit of experimental
detection.  However, the crossover from the Hamaker to the Casimir regime, 
which occurs at lengthscales of order $\nu_{\infty}^{-1}$, can be very relevant in 
practice, as it  occurs in the range of decades of nanometers. 

Adding up Eq.\ref{zero_term_main} and
Eq.\ref{third_changed_integral_main}-Eq.\ref{wide_F}, it follows that the full van der Waals forces
may be written as in Eq.(1), in terms of an effective $h$ dependent
Hamaker function, that is the sum of a constant term, corresponding to
$n=0$, and an $h$ dependent term that stems from $n>0$ contributions of
the sum in Eq.\ref{threemedsimp}. The $n=0$ term is of order $k_BT$ throughout.
At small separations, the $n>0$ term is of order $c\hbar\nu_{\infty}$, which 
corresponds to energies in the ultraviolet domain, and 
dominates largely the van der Waals interactions. At larger distances,
however,  this term vanishes altogether.
Accordingly, the $n=0$ contribution is only a small fraction
of the full interaction at small distances, but becomes increasingly more
significant as $h$ becomes large and eventually accounts for 100\% of
the  interactions in the limit $h\to \infty$.
}

\subsubsection{Analytical approximations for the four media correction}

Consistent with the linearization approximation in Eq.\ref{decomposition}, the 4-media
correction is given as:
\begin{equation}
\Delta g_{1234}(l,d) = -
\frac{k_{B}T}{2\pi}{\sum_{n=0}^{\infty}}^\prime\int_{0}^{\infty}
d\rho\,
  R(n,\rho) \, \rho\,  e^{-2(\rho_{2}l+\rho_{3}d)}
\label{first_4med}
\end{equation}
where now 
$R = \Delta_{12}^{E}\Delta_{43}^{E}  + \Delta_{12}^{M}\Delta_{43}^{M}$ 
with  $\Delta_{ij}^{E,M}$ as defined in Eq.\ref{deltas}. 

An accurate approximation to this infinite series may be obtained by
manipulating the integrand so as to adopt a form analogous to that of
the integrand in Eq.\ref{threemedsimp}. This can be achieved by
writing $\rho_2$ and $\rho_3$ in terms of the 
the auxiliary variable $\rho_{1/2}^{2} =
\rho^{2}+\frac{\epsilon_{1/2}}{c^{2}}\xi_{n}^{2}$,
with $\epsilon_{1/2}=\frac{1}{2}(\epsilon_2+\epsilon_3)$, followed
by an expansion to first order in powers of
$\Delta\epsilon=\epsilon_3-\epsilon_2$. This way, Eq.\ref{first_4med}
is transformed into:
\begin{equation}
\begin{array}{ccc}
\Delta g_{1234}(l,d)  & = & -\frac{k_{B}T}{8\pi
(l+d)^{2}}{\sum_{n=0}^{\infty}}^\prime
  \int_{r_{n}}^{\infty} dx   R^e(n,x) x e^{-x}
\end{array}
\label{beforequadrat_correc}
\end{equation}
where  $R^{e}(n,x) = R(n,x)e^{-\frac{\xi_{n}^{2}\Delta\epsilon}{c^{2}x}(d^{2}-l^{2})}$,
the lower integration limit is $r_{n} = 2(l+d)\sqrt{\epsilon_{1/2}}\frac{\xi_{n}}{c}$
and the integration variable is  $x=2\rho_{1/2}(l+d)$.

We call this the Similar Dielectric Function (SDF) approximation (a detailed development may be found in \textit{Supplementary Material 7}).
The result is now cast formally exactly as
Eq.\ref{threemedsimp} for the three media potential, so we can find approximate solutions along 
the same lines.

\paragraph{First Gaussian Quadrature Approximation}

By introducing the weight function $w(x) = x e^{-x}$, and performing a one
point Gaussian quadrature rule, we obtain an expression for $\Delta g_{1234}$
in the FGQA as a sum of a zero and finite frequency contributions:
\begin{equation}
\Delta g_{1234}^{\xi_{n} = 0}(l,d) = -\frac{k_{B}T}{16\pi (l+d)^{2}}\left(\frac{\epsilon_{1}-\epsilon_{2}}{\epsilon_{1}+\epsilon_{2}}\right)\left(\frac{\epsilon_{4}-\epsilon_{3}}{\epsilon_{4}+\epsilon_{3}}\right)
\label{zero_g1234}
\end{equation}
\begin{equation}
\Delta g_{1234}^{\xi_{n} > 0}(l,d) = -\frac{k_{B}T}{8\pi (l+d)^{2}}\sum_{n=1}^{\infty}R^{e}(n,x_{1})(1 + r_{n})e^{-r_{n}}
\label{FGQA_correction}
\end{equation}
where  $x_{1} = (2+2r_{n}+r_{n}^{2})/(1 + r_{n})$.  Equations \ref{zero_g1234} and
\ref{FGQA_correction}  correspond to the First Gaussian Quadrature Approximation
of Eq.\ref{first_4med} under the Similar Dielectric Function Approximation, and will be
designated as SDF-FGQA. Its validity will be checked in the results section.

\paragraph{Second Gaussian Quadrature Approximation}

Notice that Eq.\ref{FGQA_correction} is formally equal to Eq.\ref{rest_term_main}, so we can 
manipulate it in the same manner   as done before. Introducing the integration variable 
$\nu = \nu_{T}n = 2\sqrt{\epsilon_{1/2}}\xi_{T}n/c$, \rev{approximating}
the sum into an integral and performing
a  one point Gaussian quadrature, we obtain  (\textit{Supplementary
Material 6}):
\begin{equation}
   \Delta g_{1234}^{\xi_{n} > 0}(l+d) = -\frac{c\,\hbar \nu_{\infty}}{32\pi^{2}
(l+d)^{2} } \widetilde{R}^{e,*}_{\xi}\widetilde{F}(l+d)
\label{SGQA_corrective}
\end{equation}
where $\widetilde{R}^{e,*}_{\xi} =  \epsilon_{1/2}^{-1/2}j_{1/2}^{-1}
R^e(\nu^*,x_{1})e^{\xi}$, 
$j_{1/2} = \left( 1 + \frac{1}{2}\frac{d\ln \epsilon_{1/2}}{d \ln
\xi_{n}}\right)$ and the function $\widetilde{R}^{e,*}_{\xi}$ is 
evaluated at the quadrature point $\nu^{*}=\nu_{T} + \nu_{\infty}\xi$. Here, \rev{the factor $j_{1/2}$ will again be taken as 1 for the computation}, and
the  functions $\xi$ and $\widetilde{F}$  are as in Eq. \ref{adimensional_factor_main} and 
Eq. \ref{wide_F}, respectively, with $h=l+d$.

Since the leading order behavior of Eq.\ref{SGQA_corrective} is given by
$\widetilde{F}(l+d)$, the analogy of the correction term $\Delta g_{1234}(l+d)$
with the three media potential $g_{123}(h)$ is made obvious. In practice,
$\Delta g_{1234}^{\xi_{n} > 0}$ also depends on $l-d$, by virtue of
the factor:
\begin{equation}
   R^e(\nu^*,x_{1}) = R(\nu^*,x_{1}) e^{
   -\frac{1}{4} \frac{\Delta\epsilon}{\epsilon_{1/2}} \frac{{\nu^*}^2}{x_1}(d^2 - l^2) }
\label{R_1234_e}
\end{equation} 
However, using the results for $\nu^*$ and $x_1$,  one finds that
$R^e(\nu^*,x_{1})$ provides only corrections of order unity
to the leading order dependence provided by the function
$\widetilde{F}(l+d)$.

\subsection{Summary of results and outlook}

In this section we have provided analytical expressions for the van der Waals
free energy of two thick plates separated by two layers of variable thickness,
$l$ and $d$, $g_{1234}(l,d)$.  Starting from the exact Lifshitz result (Eq. \ref{general_lifshitz}) 
with the appropriate dispersion relation for our system (Eq. \ref{disp_rel1234}), 
we perform the expansion of the logarithm
and show that $g_{1234}(l,d)$ can be expressed in terms of
the free energies for three media, $g_{123}(l)$ and $g_{234}(d)$, together
with a correction $\Delta g_{1234}(l,d)$.

The terms $g_{123}(l)$ and $g_{234}(d)$ may be simplified by splitting the
infinite sum in $g_{123}(h)$ (Eq. \ref{threemedsimp}) into zero
(Eq.\ref{zero_term_main}) and finite (Eq.\ref{rest_term_main})
frequency contributions.  The latter may be approximated analytically by
transforming the remaining sum into an integral and applying successively 
two one-point Gaussian quadrature rules (Eq.\ref{third_changed_integral_main}).

The four media term, $g_{1234}(l,d)$ can been worked out analogously
(Eq.\ref{zero_g1234} and \ref{SGQA_corrective}) and
yields, to leading order, exactly the same distance dependence than
$g_{123}(h)$, with $h=d+l$. 

For qualitative purposes, this means that the complicated two variable
dependence of $g_{1234}^{\xi_{n} > 0}(l,d)$ can be described by a sum of 
one single variable functions, such that:
\begin{equation}
   g_{1234}^{\xi_{n} > 0}(l,d)\approx C_{123} \frac{\widetilde{F} (l)}{l^2} +
	C_{234}\frac{\widetilde{F} (d)}{d^2} + C_{1234}
	\frac{\widetilde{F}(l+d)}{(l+d)^2} 
\end{equation} 
\rev{Ignoring an order unity dependence of $C_{1234}$ on  $d^2-l^2$ that
is explicit in Eq.\ref{R_1234_e}, the 
factors $C_{ij\dots}$ may be considered  material parameters of the intervening
media. Accordingly, the leading order distance dependence of 
$g_{1234}^{\xi_{n} > 0}(l,d)$ is given by the $\widetilde{F}(h)$ functions
defined in Eq.\ref{wide_F}}.  A similar relation holds also under the 
approximation of purely pairwise additive interactions, 
with the function $\widetilde{F}(h)$
merely replaced by a constant factor \cite{mueller01,loskill12,simavilla18}.
Our approximation corrects the simplified Hamaker approximation,  predicts the
crossover from non-retarded to retarded interactions and includes corrections to
the oversimplified dependence of $\Delta g_{1234}(l,d)$ with $l+d$ that arise
when $\Delta \epsilon \ne 0$.

\section{Description of the dielectric response}

\label{sec:model}
%
So far we have described the different contributions to the surface van der Waals free energy, 
whose computation in the framework of the Lifshitz's theory requires the knowledge of the dielectric properties 
of every substance implied, essentially through the $\Delta_{ij}$ functions (Eq. \ref{deltas}) that appear in the dispersion relations. 

In order to test our theory, we need to consider an explicit model for the
dielectric properties of the system. As a simple one single component test
system, we consider the adsorption of water vapor on Silver
Iodide just below water's triple point. In this situation, we expect that a
layer of ice of arbitrary thickness $l$ can form, while, as the system
approaches the melting line, the ice surface can premelt and form a water layer
of thickness $d$. 

Since the characterization of the dielectric properties is rather cumbersome,
and the main goal of this paper is to test the theory of the previous section,
we provide here just a brief description.
A complete bibliographic review of the extinction index of these substances,
together with detailed explanation of the resulting parameterization has been presented as a part of a 
Master's thesis\citep{luengo20}, and will be published promptly in a forthcoming article.

\subsection{Dielectric response of AgI}

\label{diel}

The dielectric response of a substance evaluated at imaginary frequencies is a
real and monotonically decreasing function that drops at the frequencies at
which that material absorbs. Every valid dielectric response must fulfill the
limit $\epsilon(i\xi \rightarrow \infty) = 1$, meaning that beyond the last
absorption frequency only remains the dielectric response of the vacuum. The
dielectric function of AgI will be described following simple model called
damped oscillator\citep{hough80}, used when there are not many
experimental optical properties available in the bibliography. This
representation accounts for one absorption in the ultraviolet (UV) region and another in the infrared (IR), and the parameterization requires only the 
knowledge of the static dielectric response, $\epsilon(0)$, the refractive index
before the UV absorption, $n_{UV}$, and the absorption frequencies  $\omega_{UV}$ and $\omega_{IR}$.
\begin{equation}
\epsilon(i\xi) = 1 + \frac{\epsilon(0)-n_{UV}^{2}}{1 + (\xi/\omega_{IR})^{2}} + \frac{n_{UV}^{2}-1}{1 + (\xi/\omega_{UV})^{2}}
\label{damped_dielec}
\end{equation}
The magnitudes corresponding to the characterization of the UV absorption,
$\epsilon(0)$, $n_{UV}$ and $\omega_{IR}$, are reviewed from Bottger and
Geddes\citep{bottger67}. The remaining parameter, $\omega_{UV}$, is achieved
from a calculation based on the evolution of the refractive index published by
Cochrane\citep{cochrane74} and the Cauchy representation\citep{hough80,bergstrom97} 
(see \textit{Supplementary Material 8}). The complete parameterization is displayed in table \ref{table_paragi}.
\begin{table}[!h]
\centering
\begin{tabular}{|c|c|c|c|}
\hline 
$\epsilon(0)$ & $n_{UV}$ & $\omega_{IR}$ (eV) & $\omega_{UV}$ (eV) \\ 
\hline 
$7.00$ & $2.22$ & $1.30\ $x$\ 10^{-2}$ & $4.13$ \\ 
\hline 
\end{tabular} 
\caption{Parameterization of the dielectric response for AgI.}
\label{table_paragi}
\end{table}

\subsection{Dielectric responses of ice and liquid water}

In this work we have employed a description of the dielectric functions of
liquid water and ice achieved from the numerical fit of experimental absorption
spectra by means of the Drude model, also called Parsegian-Ninham model when it
is employed in this framework\citep{parsegian69}. The complete
bibliographic review of the extinction index of these substances, together with
the resulting parameterization, has been presented as a part of a Master's
thesis\citep{luengo20}, and will be the subject of  a forthcoming article.

For the case of water, we have selected extinction coefficients available in
the literature, and choose those measured close to 0 degrees Celsius whenever
possible \cite{zelsmann95,segelstein81,wieliczka89}.  Relative to the well known parameterization by Elbaum and Schick
\cite{elbaum91b}, our set of extinction coefficients employs 
measurements by Hayashi and Hiraoka  which provide a complete description of the high energy
band \cite{hayashi15}.
The resulting parameterization is consistent with recent work which use the
data of Ref.\citep{hayashi15} together with Infra Red absorption data at ambient temperature 
\cite{wang17,fiedler20}.

For the dielectric response of ice, there appear to be far less recent
measurements. For this reason, we have performed a parameterization largely based
on the literature review by Warren \cite{warren08}. The resulting parameterization
does not differ significantly from previous calculations by Elbaum and Schick
\cite{elbaum91b}.

As found by Fiedler et al. \cite{fiedler20}, the most significant feature in the novel
parameterization with updated experimental data by Hayashi is that the dielectric
constant of water remains higher than that of ice at all relevant finite frequencies.
As a result, the Hamaker function for the ice/water/air system is positive 
for all film thicknesses below the micron.

\section{Results and discussion}

\label{sec:results}

\subsection{Limiting cases}

\begin{figure*}[htb!]
\includegraphics[width=0.48\textwidth,height=0.35\textwidth,keepaspectratio]{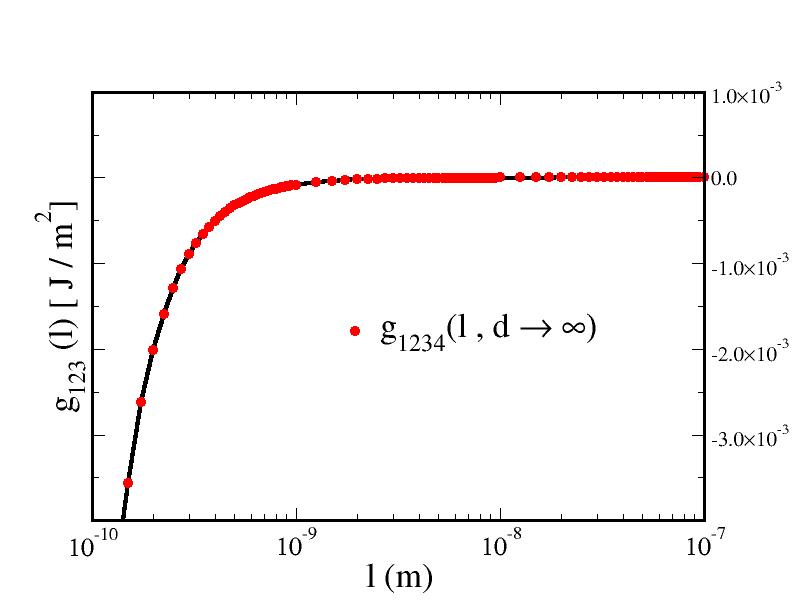}
\includegraphics[width=0.48\textwidth,height=0.35\textwidth,keepaspectratio]{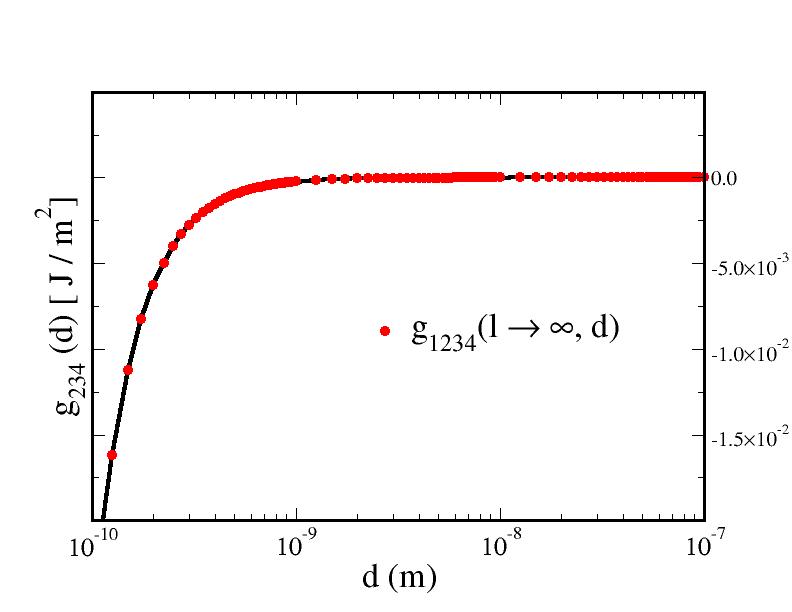}
\\
\includegraphics[width=0.48\textwidth,height=0.35\textwidth,keepaspectratio]{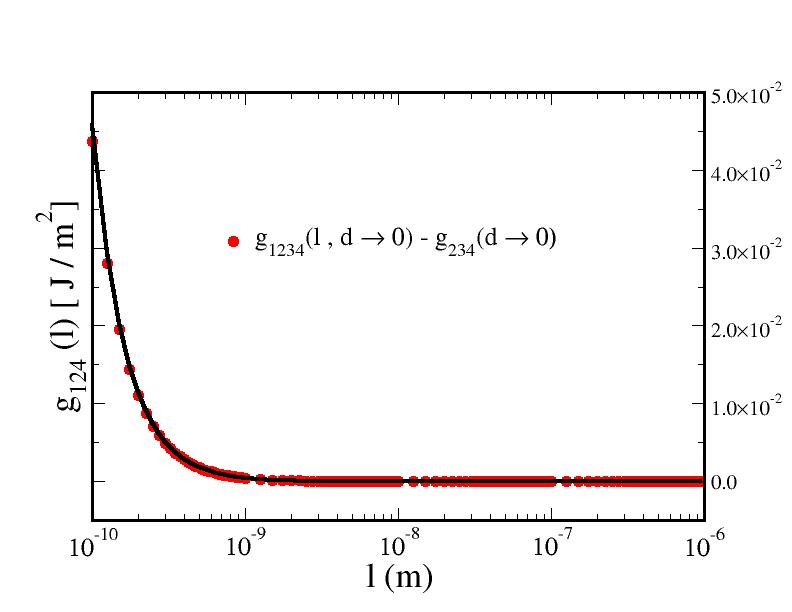}
\includegraphics[width=0.48\textwidth,height=0.35\textwidth,keepaspectratio]{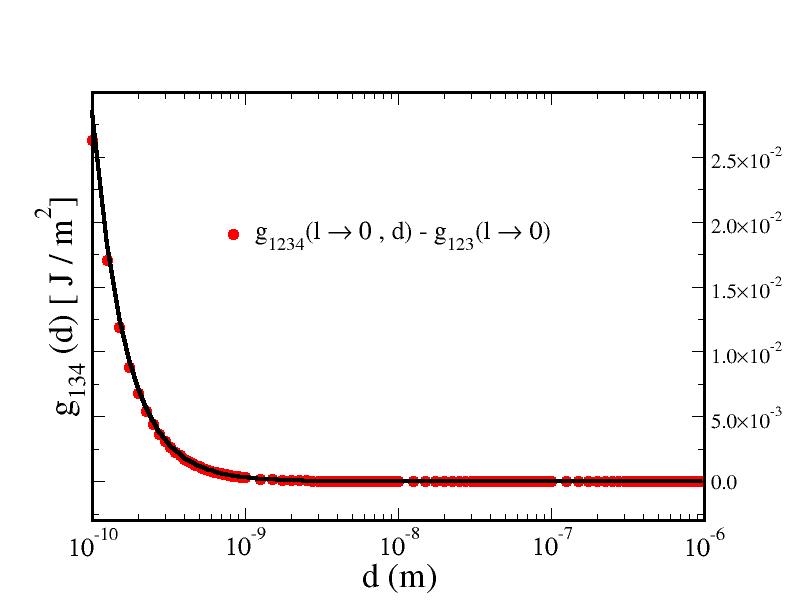}
\caption{Limiting cases of $g_{1234}(l,d)$ at infinite and zero thicknesses.  
Top left exposes the case contained in Eq. \ref{dinf}, top right represents Eq. \ref{linf}, bottom left displays the limit in 
Eq. \ref{dzero}, and bottom right the one in Eq. \ref{lzero}. Black lines
correspond to exact calculations of the \rev{interface potentials} indicated on the y
axis, while the symbols are the corresponding prediction obtained from
$g_{1234}(l,d)$ as indicated in the cited equations. The labels 1, 2, 3, 4 stand
for AgI, ice, water and vapor, respectively.
In practice for numerical purposes the large thickness limit is evaluated at
$10^{-2}$~m, while the vanishing thickness limit is evaluated at  $10^{-12}$~m.
All surface energies are given in  $J/m^{2}$.
}
\label{limiting_behaviours}
\end{figure*}

First we will check the consistency of our exact solution of $g_{1234}(l,d)$ by computing the evolution of the function at the limits of $l,d\to\infty$ and $l,d\to 0$. These correspond to the cases displayed in Eq \ref{dinf}, \ref{linf}, \ref{dzero} and \ref{lzero}, and shown in the same order in Fig \ref{limiting_behaviours}. The proper fulfillment of these special 
conditions at infinite and zero thicknesses evinces the solidity of the exact Lifshitz result and its numerical solution in the present work.
The continuity property of the \rev{interface potential} $g_{1234}(l,d)$ is very
convenient, since layers 2 and 3 need not adsorb preferentially onto the
substrate 1. Whence, in the general case where one seeks an absolute minimum of
the \rev{interface potential}, the case where either phase '2',  phase '3' or both are
not favored thermodynamically is naturally built in.

\subsection{Numerical checks}

\begin{figure*}[htb!]
\includegraphics[width=0.48\textwidth,height=0.35\textwidth,keepaspectratio]{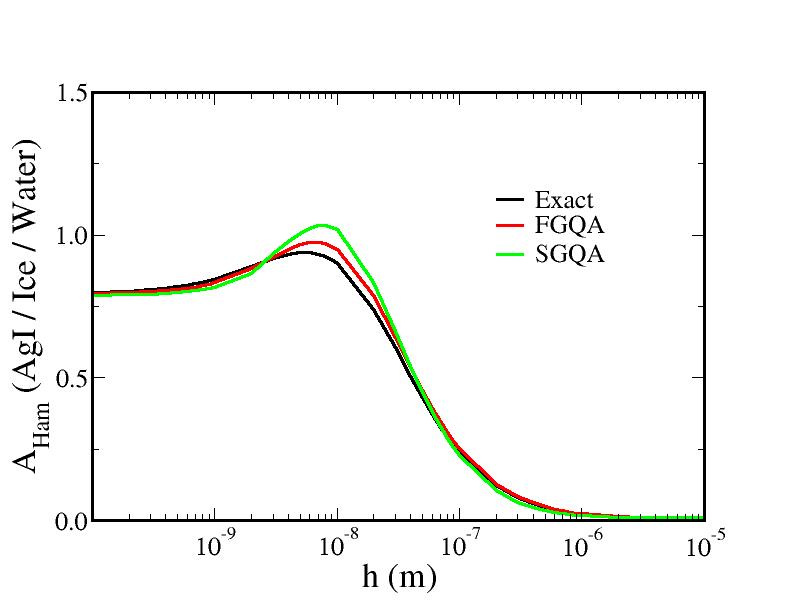}
\includegraphics[width=0.48\textwidth,height=0.35\textwidth,keepaspectratio]{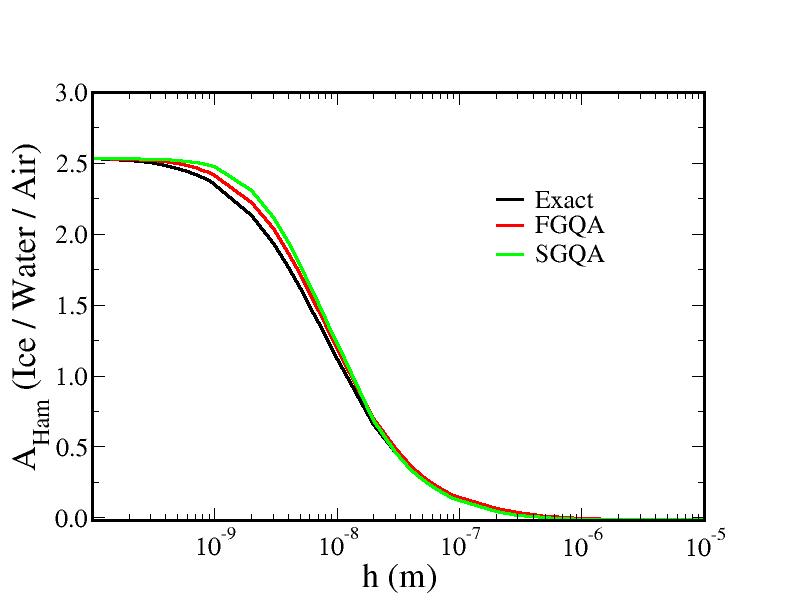}
\\
\includegraphics[width=0.48\textwidth,height=0.35\textwidth,keepaspectratio]{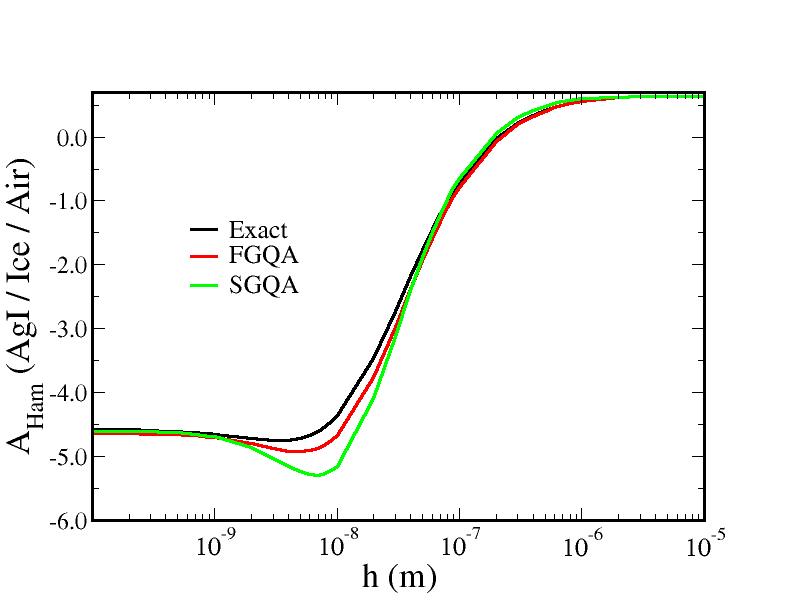}
\includegraphics[width=0.48\textwidth,height=0.35\textwidth,keepaspectratio]{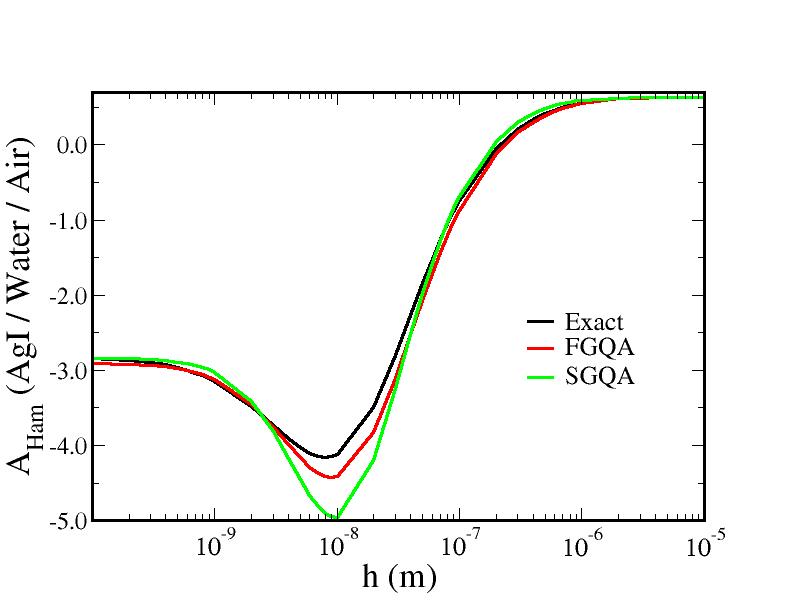}
\caption{Comparison  of exact Hamaker functions with analytical 
   approximations. \rev{The $\nu_{\infty}$ parameter for the SGQA representation is indicated below in each corresponding case.} Here we show all relevant
combinations of three media formed from AgI, ice, water and vapor: AgI/Ice/Water
(top left, \rev{$\nu_{\infty}=1.16 \times 10^{8}\ m^{-1}$}), Ice/Water/Air (top right, \rev{$\nu_{\infty}=4.10 \times 10^{8}\ m^{-1}$}), AgI/Ice/Air (bottom left, \rev{$\nu_{\infty}=1.01 \times 10^{8}\ m^{-1}$}) and
AgI/Water/Air (bottom right, \rev{$\nu_{\infty}=1.11 \times 10^{8}\ m^{-1}$}). Results are given in units of $k_{B}T$.}
\label{todos_los_3medios}
\end{figure*}

\begin{figure*}[htb!]
\includegraphics[width=0.48\textwidth,height=0.35\textwidth,keepaspectratio]{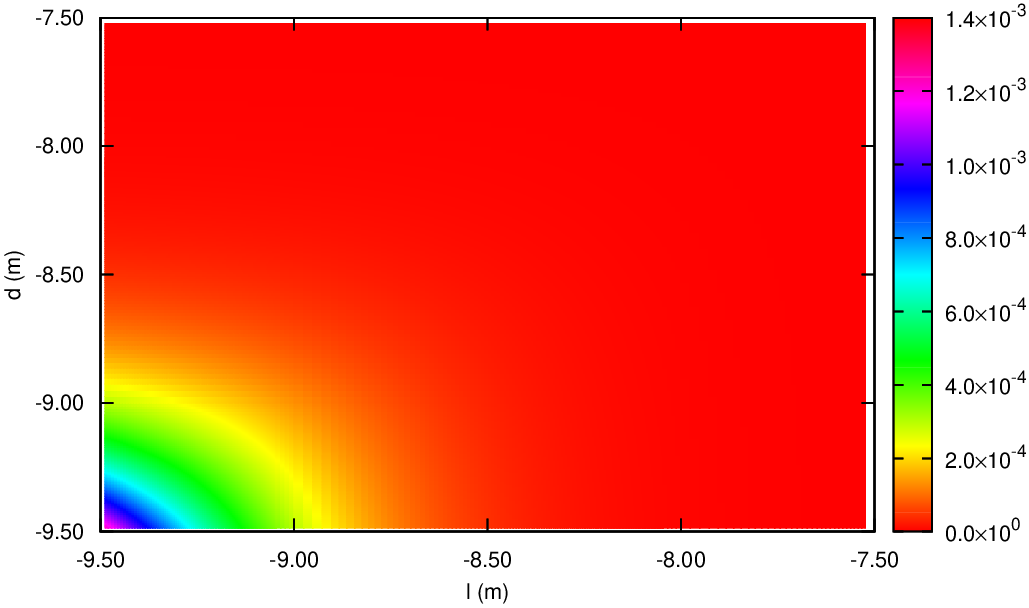}
\includegraphics[width=0.48\textwidth,height=0.35\textwidth,keepaspectratio]{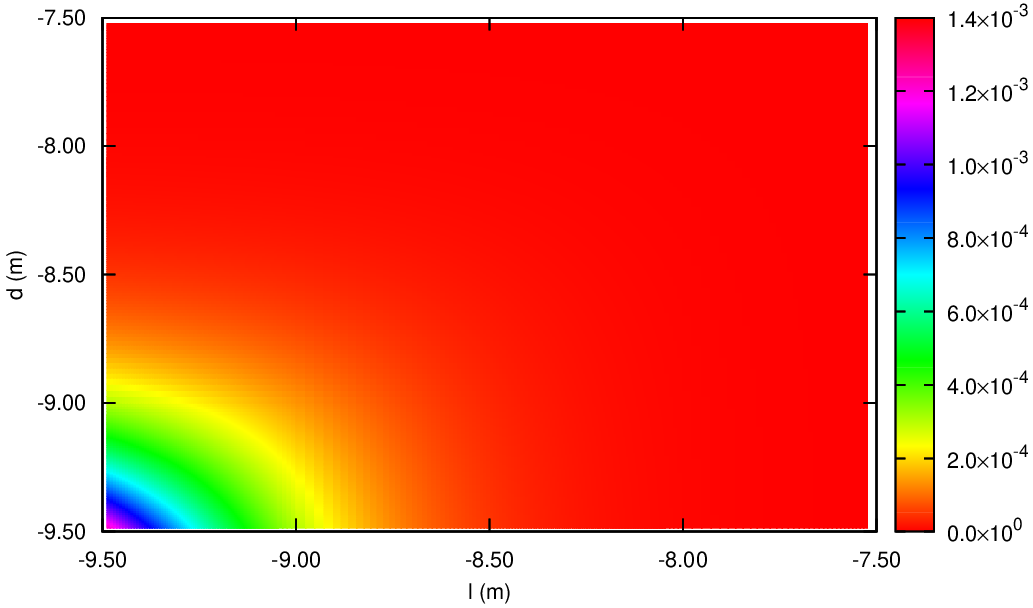}
\caption{Correction $\Delta g_{1234}(l,d)$ obtained from the exact expression
(left) and from the SDF-SGQA (right). \rev{The SFD-SGQA result is evaluated here using the parameter $\nu_{\infty}=1.06 \times 10^{8}\ m^{-1}$}. \rev{The correction} supplies a repulsive contribution when
both thicknesses ('l' for ice and 'd' for liquid water) are low, preventing the
total $g_{1234}(l,d)$ to present an absolute minimum when both the liquid water
layer and the ice layer disappear. Here the thicknesses are displayed in decimal
logarithmic scale and surface energies are given in $J/m^{2}$.}
\label{g1234_figures}
\end{figure*}

Next we demonstrate the reliability of the First Gaussian Quadrature
Approximation (FGQA) and the Second Gaussian Quadrature Approximation (SGQA) by
comparing the outcomes for the three media Hamaker function (Eq. \ref{hamfunc}).
The exact result is obtained using the general expression Eq.\ref{general_lifshitz},  
under the limiting conditions described in Eq.\ref{dinf}-\ref{linf},  while the FGQA
calculation is performed through Eq. \ref{rest_term_main} and Eq.\ref{zero_term_main}. 
The computation of the SGQA in Eq. \ref{third_changed_integral_main}
needs the knowledge of the $\nu_{\infty}$ parameter.  We achieve this by
requiring the approximate expression, Eq.\ref{third_changed_integral_main} to match the
exact free energy in the limit of vanishing film thickness.
\begin{equation}
g_{123}^{SGQA}(h\to 0) = g_{123}^{exact}(h\to 0)
\end{equation}
Essentially, this amounts to setting $\nu_{\infty}$ so as to match the exact Hamaker constant. 
\\
Figure \ref{todos_los_3medios} presents the three media Hamaker function of the systems AgI/Ice/Water, Ice/Water/Air, AgI/Ice/Air and AgI/Water/Air, and illustrates the accuracy of the analytic approximations that we have developed. The resulting Hamaker functions have been divided by $k_{B}T$ at this representation in order to make their values easier to handle, and offering also a ratio of their strength with respect to the thermal energy.
\\
With $g_{AgI/Ice/Water}$ and $g_{Ice/Water/Air}$ we have the first two terms
required to describe completely the free energy of the system (Eq.
\ref{decomposition}). The remaining contribution is given by $\Delta
g_{1234}(l,d)$. 
 We compare the exact result with the Second Gaussian Quadrature
Approximation under the Similar Dielectric Function Approximation  (SDF-SGQA)
\rev{of Eq.\ref{SGQA_corrective}} 
in Fig.\ref{g1234_figures}. \rev{As before, the computation of SDF-SGQA requires
explicit evaluation of the parameter $\nu_{\infty}$. This is achieved by
requiring that the approximate result matches the  exact value
of $\Delta g_{1234}(l,d)$ in Eq. \ref{first_4med} in the limit $l=d$ and $(l+d)\to 0$.
}
\\
\begin{figure*}[htb!]
\includegraphics[width=0.48\textwidth,height=0.35\textwidth,keepaspectratio]{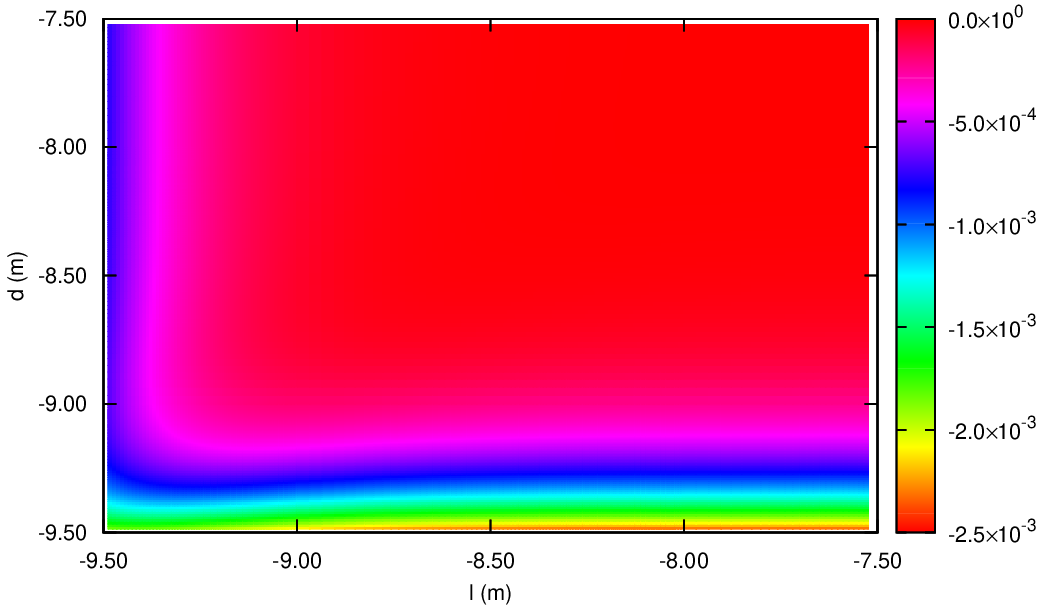}
\includegraphics[width=0.48\textwidth,height=0.35\textwidth,keepaspectratio]{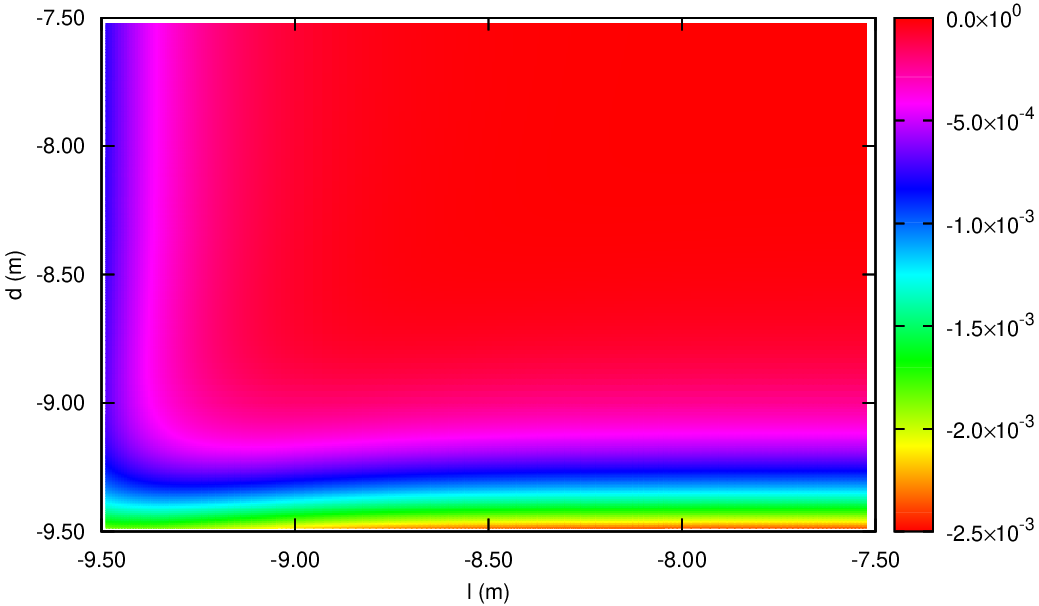}
\caption{Comparison of the full \rev{interface potential} from numerical results with
theoretical calculations. Exact $g_{1234}(l,d)$ (left), obtained from the Eq.
\ref{general_lifshitz} with the dispersion relation of Eq. \ref{disp_rel1234}.
FGQA of $g_{1234}(l,d)$ (right) under the SDF approximation. In both cases, the
axes 'l' and 'd' are the thicknesses of the ice and the liquid water layers
represented in decimal logarithmic scale, while surface free energies are given
in $J/m^{2}$.}
\label{fourmed_exact_and_app}
\end{figure*}
The numerical solution of the exact formula (Eq. \ref{general_lifshitz}) with the dispersion relation of Eq. \ref{disp_rel1234} is presented in Fig. \ref{fourmed_exact_and_app}, left.  The range is fixed from 0.3 nm to 30 nm, which is approximately the expected range of lengths of relevance of the Van der Waals forces, but the behavior is quite monotonous for wider thicknesses. This result is then taken as a reference to compare with energy curves emerging from the employment of the First Gaussian Quadrature under the Similar Dielectric Function approximation (Fig. \ref{fourmed_exact_and_app}, right), whose results arise from solving equations \ref{zero_term_main}, \ref{rest_term_main} and \ref{FGQA_correction}. The outcome never exceeds a relative error  of 3\% with respect to the exact result. Since we have proved how good this approximation works, we can now take advantage of its drastically lower computation time to perform a more exhaustive calculation of $g_{1234}(l,d)$, using now a finer mesh that otherwise would require a huge amount of time to be completed.

The FGQA approximation has also the advantage of being quite easier to compute in terms of complexity of the algorithm. The computation of the SGQA is even faster, but requires the calculation of the $\nu_{\infty}$ parameter.

\subsection{Van der Waals forces of Water and ice adsorbed on AgI}
\subsubsection{Three media systems}

In this section, we discuss the implications of every three media - Hamaker
function displayed in Fig. \ref{todos_los_3medios}. 

Recall that from Eq.
\ref{hamfunc}, the $A_{Ham}(h)$ has opposite sign to $g_{123}(h)$. Thus the
AgI/Ice/Water and Ice/Water/Air functions (top in Fig. \ref{todos_los_3medios})
provide a negative  surface free energy, with absolute minima at vanishing
film thickness. This means that  van der Waals forces do not 
favor the growth of ice at the AgI/water interface, and do not favor
the growth of a premelting film of water at the ice/air interface either.

On the other hand, both the AgI/Ice/Air and AgI/Water/Air systems (Fig. \ref{todos_los_3medios}, bottom) present negative values of the Hamaker function, 
which implies positive and monotonous decreasing surface free energies.
Accordingly, van der Waals forces favor the growth of both ice and water thick
films at the AgI/air interface. 

In practice, the ultimate behavior of growing films on the AgI surface is
dictated by a balance of short range structural forces and long range van der
Waals forces. Since the ice/air interface is known to exhibit a significant
amount of premelting \cite{slater19}, and van der Waals forces appear to
inhibit growth of a liquid film, we conclude that the existence of premelting
on ice is the result of short range structural forces, in agreement with
recent findings from computer simulation \cite{limmer14,benet19,llombart20}.

\subsubsection{AgI/Ice/Water/Air}

Once again, recall that according to Eq. \ref{decomposition}, the four media Van
der Waals free energy, $g_{1234}(l,d)$, is given as the sum of two three-media
contributions and $\Delta g_{1234}(l,d)$. The results for the four media surface
free energy, displayed in the Fig. \ref{fourmed_exact_and_app} (either left or
right), illustrate how as 'l' (the ice width) increases, the Van der Waals free
energy in the Eq. \ref{decomposition} is completely governed by $g_{234}(d)$,
and analogously, if 'd' (water layer thickness) becomes very large, the system behaves as if the air was not there, 
and the whole dispersive interaction comes now by the hand of $g_{123}(l)$. When 'l' and 'd' are negligible, 
the $\Delta g_{1234}(l,d)$ term in Eq. \ref{decomposition} contributes with a positive energy, so that the total 
$g_{1234}(l,d)$ of the system does not have an absolute minimum at $l=d=0$.
\\
Looking closely  figure \ref{fourmed_exact_and_app} at this scale, we
appreciate how the lower set of minima appears for extremely low 'd'
($d\rightarrow 0$, liquid water almost disappear) and for several - increasing
values of ice width. Indeed, the Lifshitz theory extended to four media supports
and confirms an intuitive result from the comparison between the Hamaker
functions of three media systems in the previous section: the Van der Waals
interactions favor the growth of either ice or water at the AgI/air interface
only if the  thickness of the other substance (water or ice) remains close to zero.

This is confirmed by noticing that $g_{1234}(l,d)<0$ at the bottom left corner
of Fig.\ref{fourmed_exact_and_app}. According to Eq.\ref{wettot} this implies that
the surface tension $\gamma_{AgI/air}$ is smaller than the sum of 
$\gamma_{AgI/ice}$, $\gamma_{ice/water}$ and $\gamma_{water/air}$, implying that
the adsorption of large ice and water films in between the AgI/air interface
is unfavorable.

\subsubsection{AgI/Water/Ice/Air}

\begin{figure}[!h]
\includegraphics[scale=0.23]{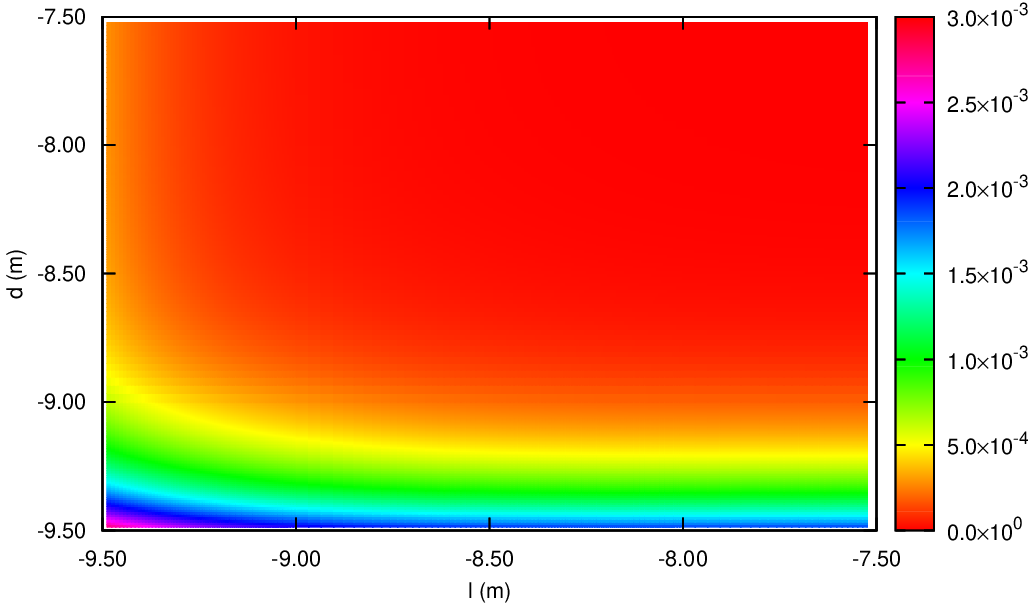}
\caption{Surface free energy for  $g_{1234}(l,d)$ and the layer order
AgI/Water/Ice/Air.  The axes 'l' and 'd' are now the thicknesses of the liquid water and the ice layers, 
in decimal logarithmic scale. Surface free energies in $J/m^{2}$.}
\label{changed_fig}
\end{figure}

We have assumed all along that the nucleation occurs with the arrangement AgI/Ice/Water/Air. Nevertheless, this is not necessarily true, and it worth to check the behavior of the system AgI/Water/Ice/Air. For this purpose we solve numerically the exact Eq. \ref{general_lifshitz} with the dispersion relation of Eq. \ref{disp_rel1234}, but this time with the following meaning of the indices: 1 = AgI, 2 = Liquid Water, 3 = Ice, 4 = Air. The result is presented in Fig. \ref{changed_fig},
\\
The resulting \rev{interface potential} is a positive (and therefore repulsive) energy,
whose zeros (minima) are placed at large values of both thicknesses, meaning
that the system will try to lower its energy by increasing the amount of both substances. Therefore, it appears that diverging layers of condensed water can grow on AgI if water is first adsorbed onto AgI and ice grows in between water and air. On the contrary, growth of ice in between AgI and water is not favored. 
The origin of this difference can be traced to the larger propensity of ice to
grow in between water and air than that of water to form in between ice and air.
This can be checked by looking at Fig. \ref{changed_fig} along the axis of large
water thickness, where it is seen that the free energy decreases by letting ice
grow. This corresponds to the phenomenon of surface freezing, which here is seen
to be favored by van der Waals forces, consistent with recent calculations
\cite{fiedler20}.

\subsubsection{Mechanism of ice nucleation}

We will use this last part of the discussion to summarize some conclusions and considerations about the role of AgI as an ice nucleator. 

Our results show that van der Waals forces promote the condensation of both ice
and water on the AgI/air interface, a result which appears to be quite in
agreement with the known nucleation efficiency of AgI \cite{marcolli16}. In
fact, we find that the Hamaker constant of the AgI/ice/air system is larger
(in absolute value) than that of AgI/water/air. This means that van der Waals
forces actually promote the freezing of water vapor over the condensation
of liquid water onto AgI. This expectation from the Hamaker constants is
confirmed by inspection of Fig.\ref{fourmed_exact_and_app}, which shows that indeed, the free
energies along the $d\to 0$ axis of vanishing water thickness are more negative
than those found along the $l\to 0$ axis of vanishing ice thickness.

On the contrary, it is found that van der Waals forces do not promote ice growth
at the AgI/water surface. This can be read off from Fig.\ref{fourmed_exact_and_app}, by looking
at the free energy along the $d\to\infty$ axis, or merely, by inspection
of the Hamaker function of the AgI/ice/water system, Fig.\ref{todos_los_3medios}. 

From these observations, we can conclude that van der Waals forces actually favor
a deposition mode of ice nucleation from the vapor phase. Of course,
this is quite at odds with experimental findings, which show that 
AgI can nucleate ice both from  vapor or water at
a similar undercooling of about 4~\textdegree{C} \cite{pruppacher10}.

The reason for this apparent discrepancy is that the ultimate behavior
of the system is dictated by a balance of both structural and van der
Waals forces. Since an undercooling of at least 4~\textdegree{C} is required
for ice to grow from supercooled water vapor, there must be short range
structural forces which oppose mildly to ice growth. Otherwise, 
in the absence of short range forces, van der Waals forces would favor
nucleation of ice from the vapor  without any undercooling.  This is
very much consistent with computer simulations by Shevkunov, which
indicate that a one monolayer thick ice film can form at ice vapor saturation,
but further growth is mildly activated \cite{shevkunov05}.

On the contrary, van der Waals forces do not promote ice growth
for AgI immersed in water, but simulations consistently show that 
Ag$^+$ exposed surfaces readily nucleate ice at mild supercooling
\cite{fraux14,zielke15,glatz16}. This
means that short range structural forces do favor ice growth at the
AgI/water surface, and a small activation is required because of the
unfavorable van der Waals interactions.

Interestingly, it has been suggested that AgI is actually most efficient
in the contact mode, whereby ice nucleation is promoted for AgI particles
in contact with condensed water droplets \cite{marcolli16}. An explanation
for this effect can be provided by assuming that nucleation actually
occurs at the three phase contact line formed between AgI, water and air 
\cite{djikaev08}, but this requires  a favorable line tension. 
Our results lend support to such favorable phenomenon. Indeed, we find that
  van der Waals forces promote
growth of a thin ice layer on AgI in contact with water vapor. Such growth
is ultimately very slow, because of the low vapor pressure of ice. However,
if such thin ice layer is formed at the contact line of an AgI particle
with the air/water interface, a large reservoir of undercooled bulk water
would be made available for the thin ice layer to continue growing
at the water/air interface, where, as we have seen, van der Waals forces
promote surface freezing.

\section{Conclusions}

\label{sec:conclusion}

In this study we work out the exact Lifshitz theory of van der Waals forces
for a substrate in the neighborhood of three phase coexistence, where
two adsorbed layers of variable thickness can form on the substrate. 
The adsorption equilibrium is set by an underlying \rev{interface potential},
as in the Frumkin-Derjaguin theory of adsorption \cite{churaev95}, but depends on the thickness of the two adsorbed layers.
Our approach goes well beyond a Hamaker theory of  pairwise additive forces
that is conventionally employed in colloidal science
\cite{hough80,bergstrom97,loskill12,simavilla18}, and extends previous results
for a single adsorption layer on substrates with coatings of fixed thickness 
\cite{ninham70c,podgornik03,parsegian05}.  

Accurate analytical approximations are provided which improve conventional 
treatments of non-retarded van der Waals forces \cite{israelachvili11} and 
apply also to the regime of retarded interactions.  This extends the validity 
of the calculations from film thicknesses barely decades of nanometers to 
arbitrary large values. Unlike the Gregory equation advocated by 
Israelachvili \cite{israelachvili11}, this is achieved without any ad-hoc 
parameter \cite{gregory81}.  By this token, we find one can now
 accurately evaluate the free energy in the Hamaker 
and Casimir regimes with the same data that is required to estimate Hamaker
constants \cite{hough80,bergstrom97,butt10,israelachvili11}.
The proper account of retardation does not only provide better accuracy. It
is a major qualitative improvement, as the crossover from non-retarded to
retarded interactions is often accompanied by a sign reversal of the
van der Waals forces.

Our results are applied to the study of water vapor adsorption on
AgI, where both layers of ice and water can form close to the triple point.
Previous studies on the ice nucleation efficiency
of AgI have provided insight into the short range interactions of
either undercooled water or vapor with the AgI surface 
\cite{shevkunov05,fraux14,zielke15,glatz16}. 
However, in the atmosphere both ice and water compete simultaneously for 
the vapor phase and it is very difficult to assess the relative stability of 
thick water and ice films from simulation.  Our results indicate
that van der Waals forces stabilize the growth of ice films at the 
AgI/air interface, but on the contrary, inhibit the growth of ice in the 
immersion mode. Importantly, van der Waals forces also promote growth of thick
ice films at the air/water interface. This explains the intriguing 
observation of sub-surface nucleation in experiments and
computer simulations \cite{durant05,haji17},
but also helps understand how the water/AgI/air contact line could promote
ice nucleation \cite{djikaev08}. The AgI/vapor surface provides the
site for stabilized ice layers that serve as embryos for the growth
of a  stable ice film at the water/vapor interface, thus lending support to  
contact mode freezing of AgI particles found in experiments \cite{marcolli16}.

Our results provide a general framework to gauge the role of van der Waals
forces in the colloidal sciences, where multicomponent solutions 
display multiphase coexistence ubiquitously, and provide a solid 
background to assess how long range forces condition the ice
nucleation efficiency of atmospheric aerosols.

\section*{Authors Contributions}

Juan Luengo-M\'arquez: Methodology, Formal analysis,  Visualization, Software,
Investigation, Writing-Original Draft;
Luis G. MacDowell: Validation, Conceptualization, Methodology,
Writing-Review and Editing.

\section*{Declaration of Competing Interests}

There are no interests to declare.

\section*{Acknowledgments}

We thank F. Izquierdo-Ruiz and Pablo Llombart  for helpful discussions and 
assistance.

\section*{Funding} 
This work was supported by the Spanish Agencia Estatal 
de Investigaci\'on under Grant No. FIS2017-89361-C3-2-P.

\section*{Data Statement}

The data used in this work is available upon request to the corresponding
author.

\section*{References}

\newpage

\title{\begin{Large}
\textbf{Supplementary Material}
\end{Large}
\vspace{0.15cm}
\\
\begin{large}
Lifshitz theory of wetting films at three phase coexistence: The case of ice nucleation
on Silver Iodide (AgI)
\end{large}
}
\date{}
\maketitle
\vspace{-2.0cm}
\begin{center}
\textit{Juan Luengo-M\'arquez and Luis G. MacDowell}
\end{center}

\subsection*{1. Derivation of the dispersion relation of the exact Lifshitz formula}

The dispersion relation of the system is an expression that all standing electromagnetic waves across the system must hold. These waves have an electric, $\vec{E}(t)$, and magnetic, $\vec{H}(t)$, field function of time, $t$, with the form
\begin{equation}
\vec{E}(t) = Re\left(\sum_{\omega}\vec{E}_{\omega}e^{-i\omega t}\right)
\label{electric_field}
\end{equation}
\begin{equation}
\vec{H}(t) = Re\left(\sum_{\omega}\vec{H}_{\omega}e^{-i\omega t}\right)
\label{magnetic_field}
\end{equation}
Being $i$ the imaginary unit, $\omega$ the frequency, and $\vec{E}_\omega$, $\vec{H}_\omega$ the amplitudes of every field at that frequency. Each must respect a certain wave equation
\begin{equation}
\nabla^{2}\vec{E} = \frac{\epsilon\mu}{c^{2}}\frac{\partial^{2}{\vec{E}}}{\partial{t}^{2}}
\label{wave_elec}
\end{equation}
\begin{equation}
\nabla^{2}\vec{H} = \frac{\epsilon\mu}{c^{2}}\frac{\partial^{2}{\vec{H}}}{\partial{t}^{2}}
\label{wave_magne}
\end{equation}
With $\epsilon$ and $\mu$ the dielectric and magnetic permeability and $c$ the velocity of light. We can turn Eq.  \ref{wave_elec} and \ref{wave_magne} into differential equations of the components of the position solving the partial derivative with time from Eq.  \ref{electric_field} and \ref{magnetic_field}, respectively. In our system, the vectors $\vec{E}$ and $\vec{H}$ have periodic x, y components with the general form $f(z)e^{i(ux+vy)}$. Then solve the derivatives in the resulting differential equation to get the general $f''(z) = \rho_{k}^{2}f(z)$, with
\begin{equation}
\rho_{k}^{2} = u^{2}+v^{2} - \frac{\epsilon_{k}\mu_{k}\omega^{2}}{c^{2}}
\end{equation}
Where we define later $\rho = u^{2}+v^{2}$. The general solution yields $f^{k}_{\alpha}(z) = A^{k}_{\alpha}e^{\rho_{k}z} + B^{k}_{\alpha}e^{-\rho_{k}z}$, with $\alpha = x, y, z$ for every substance $k$ at the system. Here we name the layers as 1/2/3/4. Finally, we impose $\nabla \cdot \vec{E}=0$ and $\nabla \cdot \vec{H}=0$ to reach that
\begin{equation}
A^{k}_{z} = -\frac{i}{\rho}(uA^{k}_{x}+vA^{k}_{y})
\label{general_A}
\end{equation}
\begin{equation}
B^{k}_{z} = \frac{i}{\rho}(uB^{k}_{x}+vA^{k}_{y})
\label{general_B}
\end{equation}
Which is true separately at every substance $k\ =$ '1', '2', '3', '4'. Next let us apply the boundary conditions at every interface, setting the $z$ axis perpendicular to the interfaces and the $z=0$ point at the interface 1/2. Now the thickness of '2' will be 'l' and that of the substance '3' will be 'd'.
\vspace{0.15cm}
\\
$\bullet$ $z = 0$, interface 1/2. Here $B^{1}_{x,y,z}$ must be zero so that $f^{1}$ does not go to infinite when z tends to $-\infty$. Then we impose the conditions $E^{1}_{x} = E^{2}_{x}$ and $E^{1}_{y} = E^{2}_{y}$, sum the resulting equations and use Eq.  \ref{general_A} and \ref{general_B} to get
\begin{equation}
-\rho_{1}A^{1}_{z} + \rho_{2}A^{2}_{z} - \rho_{2}B^{2}_{z} = 0
\label{1}
\end{equation}
On the other hand, it must also be true that $\epsilon_{1}E^{1}_{z} = \epsilon_{2}E^{2}_{z}$, that is
\begin{equation}
\epsilon_{1}A^{1}_{z} = \epsilon_{2}A^{2}_{z}+\epsilon_{2}B^{2}_{z}
\label{2}
\end{equation}
\vspace{0.15cm}
\\
$\bullet$ $z = l$, interface 2/3. Here we apply basically the same procedure. Notice that now the exponentials do not vanish, so we reach from $E^{2}_{x} = E^{3}_{x}$ and $E^{2}_{y} = E^{3}_{y}$
\begin{equation}
-\rho_{2}A^{2}_{z}e^{\rho_{2}l} + \rho_{2}B^{2}_{z}e^{-\rho_{2}l} + \rho_{3}A^{3}_{z}e^{\rho_{3}l}-\rho_{3}B^{3}_{z}e^{-\rho_{3}l} = 0
\label{3}
\end{equation}
And from the condition $\epsilon_{2}E^{2}_{z} = \epsilon_{3}E^{3}_{z}$
\begin{equation}
\epsilon_{2}A^{2}_{z}e^{\rho_{2}l} + \epsilon_{2}B^{2}_{z}e^{-\rho_{2}l} = \epsilon_{3}A^{3}_{z}e^{\rho_{3}l} + \epsilon_{3}B^{3}_{z}e^{-\rho_{3}l}
\label{4}
\end{equation}
\vspace{0.15cm}
\\
$\bullet$ $z = l+d$, interface 3/4. Here $A^{4}_{x,y,z}$ must be zero so that $f^{4}$ does not go to infinite as z tends to $\infty$. From $E^{3}_{x} = E^{4}_{x}$ and $E^{3}_{y} = E^{4}_{y}$ now we have
\begin{equation}
-\rho_{3}A^{3}_{z}e^{\rho_{3}(l+d)} + \rho_{3}B^{3}_{z}e^{-\rho_{3}(l+d)} - \rho_{4}B^{4}_{z}e^{-\rho_{4}(l+d)} = 0
\label{5}
\end{equation}
And finally, from $\epsilon_{3}E^{3}_{z} = \epsilon_{4}E^{4}_{z}$ we get
\begin{equation}
\epsilon_{3}A^{3}_{z}e^{\rho_{3}(l+d)} + \epsilon_{3}B^{3}_{z}e^{-\rho_{3}(l+d)} = \epsilon_{4}B^{4}_{z}e^{-\rho_{4}(l+d)}
\label{6}
\end{equation}
So far we have used the boundaries for the electric fields. Solving the system
of equations formed by Eq.  \ref{1}, \ref{2}, \ref{3}, \ref{4}, \ref{5} and
\ref{6} for the six variables $\lbrace A_{i}\rbrace$, $\lbrace B_{i}\rbrace$,
leads to \rev{what equalized to zero is the dispersion relation of the system $D_{M}=0$}. For that we just arrange those equations in the determinant
\vspace{0.15cm}
\\
\begin{center}
$\left|\begin{array}{cccccc}
-\rho_{1} & \rho_{2} & -\rho_{2} & 0 & 0 & 0 \\ 
\epsilon_{1} & -\epsilon_{2} & -\epsilon_{2} & 0 & 0 & 0 \\ 
0 & -\rho_{2}e^{\rho_{2}l} & \rho_{2}e^{-\rho_{2}l} & \rho_{3}e^{\rho_{3}l} & -\rho_{3}e^{-\rho_{3}l} & 0 \\ 
0 & \epsilon_{2}e^{\rho_{2}l} & \epsilon_{2}e^{-\rho_{2}l} & -\epsilon_{3}e^{\rho_{3}l} & -\epsilon_{3}e^{-\rho_{3}l} & 0 \\ 
0 & 0 & 0 & -\rho_{3}e^{\rho_{3}(l+d)} & \rho_{3}e^{-\rho_{3}(l+d)} & -\rho_{4}e^{-\rho_{4}(l+d)} \\ 
0 & 0 & 0 & \epsilon_{3}e^{\rho_{3}(l+d)} & \epsilon_{3}e^{-\rho_{3}(l+d)} & -\epsilon_{4}e^{-\rho_{4}(l+d)}
\end{array}\right|=D_{M}$
\end{center}

Solving for the determinant explicitly gives a sum of 32 terms which is not
particularly insightful. However, one notices that products of two matrix
elements of the form $d_{ij} d_{kl}$ have common factors with terms $d_{il}
d_{kj}$, or alternatively, terms of the form $\rho_i\epsilon_j$ share common
factors with $\rho_j\epsilon_i$. Therefore, we organize the 32 terms as:
\begin{equation}
   \begin{array}{c}
	  D_M = \\
    (\rho_1\epsilon_2 - \rho_2\epsilon_1)
   \left [  (\rho_4\epsilon_3 - \rho_3 \epsilon_4 ) ( \rho_2 \epsilon_3 + \rho_3 \epsilon_2 )
	\alpha^{-1}\beta\gamma^{-1}   + 
	      (\rho_4 \epsilon_3 + \rho_3 \epsilon_4) ( \rho_3 \epsilon_2 - \rho_2
		\epsilon_3 ) \alpha^{-1}\beta^{-1}\gamma
   \right ] \delta^{-1} \\
	    + \\
	   (\rho_2 \epsilon_1 + \rho_1 \epsilon_2)
	   \left [ (\rho_2 \epsilon_3 - \rho_3 \epsilon_2 )(\rho_3 \epsilon_4 - \rho_4 \epsilon_3)\alpha\beta\gamma^{-1} -
		(\rho_3 \epsilon_2 + \rho_2 \epsilon_3 ) ( \rho_4 \epsilon_3 + \rho_3 \epsilon_4 ) \alpha\beta^{-1}
		 \gamma
   \right ] \delta^{-1}
\end{array}
\end{equation} 
where we have introduced the symbols $\alpha=e^{\rho_2 l}$, $\beta=e^{\rho_3
l}$, $\gamma = e^{\rho_3(l+d)}$ and $\delta=e^{\rho_4(l+d) }$, for short.

Notice now that the determinant is a product of elements $(\rho_i \epsilon_j +
\rho_j\epsilon_i)$, which correspond to denominators of the function
$\Delta_{ij}^M$ in the main text (see also below, Eq.\ref{delta_M}); 
and elements $(\rho_i \epsilon_j - \rho_j\epsilon_i)$ which correspond to numerators of $\Delta_{ij}^M$.
Since, from the dispersion relation, this determinant must vanish, we can now
multiply and divide by constants without changing the result. Therefore,
we divide by factors of the form $(\rho_i \epsilon_j +
\rho_j\epsilon_i)$, and further divide by $\alpha\beta^{-1}\gamma\delta^{-1}$ to get:
\begin{equation}
  D_M =
1 - \Delta^{M}_{12}\Delta^{M}_{32}e^{-2\rho_{2}l} - \Delta^{M}_{23}\Delta^{M}_{43}e^{-2\rho_{3}d} - \Delta^{M}_{12}\Delta^{M}_{43}e^{-2\rho_{2}l}e^{-2\rho_{3}d} 
\label{disp_rel_M}
\end{equation}
\begin{equation}
\Delta^{M}_{ij} = \frac{\rho_{j}\epsilon_{i}-\rho_{i}\epsilon_{j}}{\rho_{j}\epsilon_{i}+\rho_{i}\epsilon_{j}}
\label{delta_M}
\end{equation}
which is the sought result. Notice that here $l$ and $d$ stand on the
same footing.

The result for a system of one single medium between two plates, with one
plate coated by a layer of fixed thickness, $d$, can be obtained from this
expression readily. In that situation  the physical
requirement is that $D_M\to 1$, as $l\to\infty$ at fixed $d$. From
Eq.\ref{disp_rel_M}, we find instead:
\begin{equation}
   \lim_{l\to\infty} D_M = 1 - \Delta_{23}^M\Delta_{43}^M e^{-2\rho_3 d}
\end{equation} 
Therefore, a new \rev{function} consistent with the mentioned
physical requirement can be obtained simply dividing $D_M$ by
$\lim_{l\to\infty}  D_M$. 
This yields:
\begin{equation}
   D'_M = 1 - \frac{\Delta_{12}^M\Delta_{32}^M+\Delta_{12}^M\Delta_{43}^Me^{-2\rho_3
   d}}{1- \Delta_{23}^M\Delta_{43}^M e^{-2\rho_3 d} } e^{-2\rho_2 l}
\end{equation} 
which is the well known \rev{relation} used for systems with one
coated layer of fixed size. Because of the
choice of normalization condition, 
the variables $l$ and $d$ no longer stand on the same footing.

\rev{
The calculation of
$D_{E}$ arises from a similar analysis of the boundaries 
for the magnetic fields, and takes the same form as Eq.(\ref{disp_rel_M}), with
$\Delta_{ij}^M$ merely replaced by $\Delta_{ij}^E$,
\begin{equation}
\Delta^{E}_{ij} = \frac{\rho_{j}-\rho_{i}}{\rho_{j}+\rho_{i}}
\label{delta_E}
\end{equation}
With $D_E$ and $D_M$ at hand, we readily obtain $D=D_{M}\cdot D_{E}$,
which can be plugged into the general expression for the 
interface potential, Eq.(6) of the main text.
Alternatively, we can transform $D_E$ to obtain $D'_E$ in the
same way as we obtained $D'_M$, which yields $D'=D'_E\cdot D'_M$.

The significance of this transformation may be understood by
noticing that we have defined in practice:
\begin{equation}
        D'_{1234}(l;d) = D_{1234}(l,d) \left / D_{1234}(l\to\infty,d) \right .
\end{equation} 
where the prime next to D and the
 arguments $(l;d)$ in the left hand side indicate that
the thickness 'd' is now meant as a fixed parameter, not a variable.

Plugging this result into Eq.(6) of the main text, we readily find:
\begin{equation}
   g'_{1234}(l;d) = \frac{k_{B}T}{2
\pi}{\sum_{n=0}^{\infty}}^\prime\int_{0}^{\infty}\rho
\hspace{0.1cm} d\rho \ln
        \left(\frac{D_{1234}(l,d)}{D_{1234}(l\to\infty,d)}\right)
\end{equation}
By splitting the logarithm into two different additive contributions,
we readily find:
\begin{equation}
      g'_{1234}(l;d) = g_{1234}(l,d) - g_{1234}(l\to\infty,d)
\end{equation} 
Whence, the known results for the interface potential of a system
with layer '2' of variable thickness $l$ and a coating of fixed thickness $d$
obtained previously in Ref.26--28.
differs from the full two variable
interface potential merely by a normalization constant, which
sets the requirement that $g'_{1234}(l;d)$ vanishes in the limit
that $l\to\infty$.
Alternatively, noticing
that in fact $g_{1234}(l\to\infty,d)$ corresponds to
the interface potential for a three media system of layer '3' sandwiched
between semi-infinite bodies '2' and '4' (c.f. Eq.(10) of main text), 
the above result may be readily written as:
\begin{equation}
      g'_{1234}(l;d) = g_{1234}(l,d) - g_{234}(d)
\end{equation} 
Particularly, in the limit that $d$ vanishes altogether, $g'_{1234}(l;d\to 0)$
becomes $g_{124}(l)$, so we find:
\begin{equation}
      g_{124}(l) = g_{1234}(l,d\to 0) - g_{234}(d\to 0)
\end{equation} 
which is Eq.(11) of the main text. 
A similar analysis holds for the case $l\to 0$, which corresponds to
Eq.(12) of the main text.

}

\subsection*{2. Change of variable to 'x'}
Our purpose here is to develop explicitly the route from the three media exact
surface free energy to the equation over which the First Gaussian Quadrature
Approximation is performed,
Eq.14 of the main text. Then we start from the generalized form of the three media surface potential
\begin{equation}
   g_{123}(h) = \frac{k_{B}T}{2\pi}{\sum^{\infty}_{n=0}}^\prime\int_{0}^{\infty}\rho \hspace{0.05cm}d\rho \ln(D^{E}_{123}D^{M}_{123})
\label{eq6_article}
\end{equation}
Where '$h$' stands for the thickness of the layer '$2$' and
\rev{the prime next to the sum indicates that the term $n=0$ has an
extra factor $1/2$.}
The dispersion relation
is that of the three layered media geometry, as dictated by $D^{E,M}_{123}=1-\Delta^{E,M}_{12}\Delta^{E,M}_{32}e^{-2\rho_{2}h}$. Next we assume that the term after the 1 at $D^{E,M}_{123}$ is small, and we expand the logarithm as $ln(1-x)\approx=-x$. At the same time, let us change the variable to $\rho_{2}^{2}=\rho^{2}+\frac{\epsilon_{2}\xi_{n}^{2}}{c^{2}}$. We perform this transformation simply through $\rho_{2}d\rho_{2} = \rho d\rho$, and the lower limit of the integral now is $\rho_{2}(\rho =0)=\sqrt{\frac{\epsilon_{2}\xi_{n}^{2}}{c^{2}}}$
\begin{equation}
   g_{123}(h) = -\frac{k_{B}T}{2\pi}{\sum^{\infty}_{n=0}}^\prime\int_{\sqrt{\frac{\epsilon_{2}\xi_{n}^{2}}{c^{2}}}}^{\infty}\rho_{2} \hspace{0.05cm}d\rho_{2} (\Delta^{M}_{12}\Delta^{M}_{32} + \Delta^{E}_{12}\Delta^{E}_{32})e^{-2\rho_{2}h}
\label{firstchange}
\end{equation}
With $\Delta^{M}_{ij}$ and $\Delta^{E}_{ij}$ as defined in Eq. \ref{delta_M}
and Eq.\ref{delta_E}.

In the last step we perform a second change of variables to $x = 2\rho_{2}h$, so that $x \hspace{0.05cm}dx = 4h^{2}\rho_{2} \hspace{0.05cm}d\rho_{2}$. Moreover, the lower limit is transformed again to $2h\sqrt{\frac{\epsilon_{2}\xi_{n}^{2}}{c^{2}}}$, which we define as $r_{n}$
\begin{equation}
g_{123}(h) = -\frac{k_{B}T}{8\pi h^{2}}{\sum^{\infty}_{n=0}}^\prime\int_{r_{n}}^{\infty}x \hspace{0.05cm}dx (\Delta^{M}_{12}\Delta^{M}_{32} + \Delta^{E}_{12}\Delta^{E}_{32})e^{-x}
\label{eq7_article}
\end{equation}
Both $\Delta^{M}_{ij}$ and $\Delta^{E}_{ij}$ may be expressed in terms of $x_{i} = \sqrt{x^{2} + (\epsilon_{i}-\epsilon_{2})(2h\xi_{n}/c)^{2}}$ instead of $\rho_{i}$, through the straightforward substitution $x_{i}=2\rho_{i}h$. 
The Eq.   \ref{eq7_article} corresponds to the stage immediately before the FGQA at the main part of the work, and is the result to which we wanted to arrive here.

\subsection*{3. Gaussian Quadrature}
The Gaussian quadrature is an integration method where the integrand is separated between a well shaped function, $f(x)$, and a weight function, $w(x)$. The generalized N points Gaussian quadrature reads
\begin{equation}
\int_{a}^{b}f(x)w(x)dx = \sum_{i=1}^{N}f(x_{i})m_{i}
\label{Ngaussian}
\end{equation}
With $f(x)$ evaluated at the nodes $x_{i}$ (also called quadrature points). The knowledge of the sets of $\lbrace x_{i}\rbrace$ and $\lbrace m_{i}\rbrace$ requires the capability of solving from the $j=0$ up to the $2N-1$ integral of the kind
\begin{equation}
I_{j} = \int_{a}^{b}x^{j}w(x)dx = \sum_{i=1}^{N}x_{i}^{j}m_{i}
\label{integral_quadrar}
\end{equation}
That leads to a system of 2N equations whose solution provides the quadrature points and all $\lbrace m_{i}\rbrace$. In this work we have used the one point Gaussian quadrature (N=1), so the integrals that we have to solve are
\begin{equation}
I_{0} = \int_{a}^{b}w(x)dx
\label{integral1}
\end{equation}
\begin{equation}
I_{1} = \int_{a}^{b}x \hspace{0.05cm} w(x)dx
\label{integral2}
\end{equation}
Thus we have $m_{1}=I_{0}$, and $x_{1}=I_{1}/I_{0}$.

\subsection*{4. One point Gaussian quadrature applied to the FGQA}
We wish to specify here the quadrature performed over
\begin{equation}
g^{\xi_{n}>0}(h) = -\frac{k_{B}T}{8\pi h^{2}}\sum^{\infty}_{n=1}\int_{r_{n}}^{\infty}x \hspace{0.05cm}dx R(n,x)e^{-x}
\end{equation}
which is written here in generalized form so that it applies both for
$g_{123}(h)$ and $\Delta g_{1234}(l,d)$. In the last case just consider that $h
= (l+d)$, and employ $R^{e}(n,x)$ instead of $R(n,x)$. We state then $f(x) =
R(n,x)$, and $w(x) = xe^{-x}$. Using this weight in
Eq.\ref{integral1}-\ref{integral2} provides straightforward integrals of the form
\begin{equation}
I_{0}=\int_{r_{n}}^{\infty}xe^{-x}dx = e^{-r_{n}}(1+r_{n})
\end{equation}
\begin{equation}
I_{1}=\int_{r_{n}}^{\infty}x^{2}e^{-x}dx = e^{-r_{n}}(2+2r_{n}+r_{n}^{2})
\end{equation}

\subsection*{5. Euler - MacLaurin formula}
The Euler-MacLaurin formula allows the transformation of a summation into an integral through an approximation, and reads
\begin{equation}
\sum_{n=a}^{b}f(n) = \int_{a}^{b}f(n)dn + \frac{1}{2}(f(a)+f(b))+\sum^{\infty}_{k=1} \frac{B_{2k}}{2k!}(f^{(2k-1)}(b)-f^{(2k-1)}(a))
\label{euler_maclaurin}
\end{equation}
$B_{i}$ \rev{being} the Bernoulli coefficients and \rev{$f^{(m)}$} the 
m-th derivative of $f$. From now on we will apply it up to the first corrective order, $k=1$.
\vspace{0.15cm}
\\
We start again from the generalization presented in the previous section. We have after the FGQA the expression
\begin{equation}
g^{\xi_{n}>0}(h) = -\frac{k_{B}T}{8\pi h^{2}}\sum^{\infty}_{n=1}R(n,x_{1})(1+r_{n})e^{-r_{n}}
\end{equation}
And now we evaluate the corrective term of the Eq. \ref{euler_maclaurin}, considering that $R(n,x_{1})$ is essentially constant with $n$ compared to the exponential dependence. This is also true for the $\Delta g_{1234}(l,d)$ particularization, since even if the exponential in $R^{e}(n,x_{1})$ contains a $n^{2}$ factor, the value of $\Delta\epsilon$ is extremely close to zero for large $n$, and the value of $(d^{2}-l^{2})$ is quite small as well. These features make that exponential factor very close to one. The corrective term of the Euler-MacLaurin formula is then
\begin{equation}
\Delta g^{\xi_{n}>0} = -\frac{k_{B}T}{8\pi h^{2}}\left[\frac{1}{2}R(1,x_{1})(1+r_{T})e^{-r_{T}} + \frac{1}{12}R(1,x_{1})r_{T}^{2}e^{-r_{T}}\right]
\end{equation}
Where $r_{T}=r_{n=1}$. For most cases of interest, where $\nu_T h \ll 1$, this
term is negligible and adds a correction of order $\nu_T/\nu_{\infty}$ to the
Hamaker constant. In the regime where $\nu_T h \approx 1$, it exhibits the same
order of magnitude as the leading order term, and does therefore not upset the 
scaling. However, in that range the finite frequency contribution becomes
negligible compared to the $n=0$ term, so it is save to neglect it altogether.
Accordingly, we neglect the corrections here for the sake of simplicity and
use the approximation: 
\begin{equation}
g^{\xi_{n}>0}(h) = -\frac{k_{B}T}{8\pi h^{2}}\int^{\infty}_{1}R(n,x_{1})(1+r_{n})e^{-r_{n}}dn
\end{equation}
The following step is to change the variable to $\nu = (4\pi k_{B}T \epsilon^{1/2}_{2}n)/(c\hbar)$. Realize that $\epsilon_{2}$ is also a function of n, so we change the variable through
\begin{equation}
d\nu = \frac{4\pi k_{B}T}{c\hbar}\epsilon^{1/2}_{2}dn\left[1 + \frac{1}{2}\frac{d\ln\epsilon_{2}}{d\ln\xi_{n}}\right]
\label{change_variable}
\end{equation}
The term inside the brackets is defined as $j_{2}$, and its value is approximately 1. If the last transformations are to be performed at $\Delta g_{1234}(l,d)$, we employ $\epsilon_{1/2}$ instead of $\epsilon_{2}$, and correspondingly, we define $j_{1/2}$ instead of $j_{2}$. Once the variable is changed we achieve
\begin{equation}
g^{\xi_{n}>0}(h) = -\frac{c\hbar}{32\pi^{2} h^{2}}\int^{\infty}_{\nu_{T}}\widetilde{R}(\nu,x_{1})(1+h\nu)e^{-h\nu}d\nu
\label{citecite}
\end{equation}
Where $\widetilde{R}(\nu,x_{1})= \epsilon_{2}^{-1/2}j_{2}^{-1}R(\nu,x_{1})$.

\subsection*{6. One point Gaussian quadrature applied to the SGQA}
Beginning at the Eq.   \ref{citecite}, we introduce the auxiliary exponential function $e^{-\nu/\nu_{\infty}}$, with the parameter $\nu_{\infty}$ chosen to mimic the scale at which the algebraic decay of $\widetilde{R}(\nu,x_{1})$ becomes significant
\begin{equation}
g^{\xi_{n}>0}(h) = -\frac{c\hbar}{32\pi^{2} h^{2}}\int^{\infty}_{\nu_{T}}\widetilde{R}(\nu,x_{1})e^{\nu/\nu_{\infty}}\left[ e^{-\nu/\nu_{\infty}} (1+h\nu)e^{-h\nu}\right] d\nu
\end{equation}
And next we perform the one point Gaussian quadrature approximation with $f(\nu) =  \widetilde{R}(\nu,x_{1})e^{\nu/\nu_{\infty}}$, and $w(\nu)=e^{-\nu/\nu_{\infty}}(1+h\nu)e^{-h\nu}$. We solve then $I_{0}$ and $I_{1}$ by parts
\begin{equation}
I_{0} = \int_{\nu_{T}}^{\infty}e^{-\nu/\nu_{\infty}}(1+h\nu)e^{-h\nu}d\nu
\end{equation}
\begin{equation}
I_{0} = \nu_{\infty}\frac{(\nu_{T}h+1)(\nu_{\infty}h+1)+\nu_{\infty}h}{(\nu_{\infty}h+1)^{2}}e^{-\nu_{T}h-\frac{\nu_{T}}{\nu_{\infty}}}
\end{equation}
\begin{equation}
I_{1} = \int_{\nu_{T}}^{\infty}\nu e^{-\nu/\nu_{\infty}}(1+h\nu)e^{-h\nu}d\nu
\end{equation}
\begin{equation}
I_{1} = \nu_{\infty}\frac{(\nu_{T}h+1)(\nu_{\infty}h+1)^{2}\nu_{T}+  (2\nu_{T}h+1)(\nu_{\infty}h+1)\nu_{\infty}  + 2\nu_{\infty}^{2}h}{(\nu_{\infty}h+1)^{3}}e^{-\nu_{T}h-\frac{\nu_{T}}{\nu_{\infty}}}
\end{equation}

\subsection*{7. Similar Dielectric Function approximation}
In this section we intend to clarify the steps to follow throughout the development of the Similar Dielectric Function approximation. Starting from the correction term of the four media surface free energy
\begin{equation}
   \Delta g_{1234}(l,d) = -\frac{k_{B}T}{2\pi}{\sum_{n=0}^{\infty}}^\prime\int_{0}^{\infty}\rho \ d\rho \ R_{1234}(n,\rho) \  e^{-2(\rho_{2}l+\rho_{3}d)}
\label{first_4med_222}
\end{equation}
We first notice from the definition of $\rho_i$ that one can write
\begin{equation}
\rho_{2}^{2}=\rho_{1/2}^{2}-\frac{1}{2}\frac{\Delta\epsilon}{c^{2}}\xi_{n}^{2}
\label{rho2_basic}
\end{equation}
\begin{equation}
\rho_{3}^{2}=\rho_{1/2}^{2}+\frac{1}{2}\frac{\Delta\epsilon}{c^{2}}\xi_{n}^{2}
\label{rho3_basic}
\end{equation}
Where we have introduced $\Delta\epsilon = \epsilon_{3}-\epsilon_{2}$, and $\rho_{1/2}^{2} = \rho^{2}+\frac{1}{2}\frac{(\epsilon_{3}+\epsilon_{2})}{c^{2}}\xi_{n}^{2}$. Then we assume small $\Delta\epsilon$ and apply Taylor to get
\begin{equation}
\rho_{2} \approx \rho_{1/2} - \frac{1}{4}\frac{\xi_{n}^{2}\Delta\epsilon}{c^{2}\rho_{1/2}}
\label{expansion_rho2}
\end{equation}
\begin{equation}
\rho_{3} \approx \rho_{1/2} + \frac{1}{4}\frac{\xi_{n}^{2}\Delta\epsilon}{c^{2}\rho_{1/2}}
\label{expansion_rho3}
\end{equation}
Using these results, the exponential function in Eq.\ref{first_4med_222} can be 
now factored into two simpler exponentials:
\begin{equation}
e^{-2(\rho_{2}l+\rho_{3}d)} \approx e^{-2\rho_{1/2}(l+d)}e^{-\frac{1}{2}\frac{\xi_{n}^{2}\Delta\epsilon}{c^{2}\rho_{1/2}}(d-l)}
\label{exponential_development}
\end{equation}
Eq. \ref{exponential_development} is the cornerstone of the Similar Dielectric Function approximation. Replacing in Eq. \ref{first_4med_222}, we can now express $\Delta g_{1234}(l,d)$ as
\begin{equation}
   \Delta g_{1234}(l,d)  =  -\frac{k_{B}T}{2\pi}{\sum_{n=0}^{\infty}}^\prime
\int_{\sqrt{\epsilon_{1/2}}\frac{\xi_{n}}{c}}^{\infty} d\rho_{1/2} \rho_{1/2}R(n,\rho_{1/2})e^{-2\rho_{1/2}(l+d)}e^{-\frac{1}{2}\frac{\xi_{n}^{2}\Delta\epsilon}{c^{2}\rho_{1/2}}(d-l)}
\label{main_4med_222}
\end{equation}
Where we have switched the integration variable to $\rho_{1/2}$ 
and we have defined the mean dielectric response of the intervening media  as:
\begin{equation}
\epsilon_{1/2} = \frac{1}{2}(\epsilon_{2}+\epsilon_{3})
\label{epsi_unmedio}
\end{equation}
With the purpose of following an analogous path to that previously described 
for the three media terms, we perform a second change of variable under the
definition $x=2\rho_{1/2}(l+d)$, leading to
\begin{equation}
\begin{array}{ccc}
   \Delta g_{1234}(l,d)  & = & -\frac{k_{B}T}{8\pi
   (l+d)^{2}}{\sum_{n=0}^{\infty} }^\prime
  \int_{r_{n}}^{\infty} dx \ x \  R^e(n,x) e^{-x}
\end{array}
\label{beforequadrat_correc}
\end{equation}
Under the definitions $R^{e}(n,x) = R(n,x)e^{-\frac{\xi_{n}^{2}\Delta\epsilon}{c^{2}x}(d^{2}-l^{2})}$,
and $r_{n} = 2(l+d)\sqrt{\epsilon_{1/2}}\frac{\xi_{n}}{c}$. This expression is now very similar to the three media potential and can be treated through the same approximations.

\subsection*{8. Parameterization of the damped oscillator model for AgI}

The damped oscillator model employed for the AgI describes the dielectric function at imaginary frequencies as
\begin{equation}
\epsilon(i\xi) = 1 + \frac{\epsilon(0)-n^{2}_{UV}}{1+(\xi/\omega_{IR})^{2}} + \frac{n^{2}_{UV}-1}{1+(\xi/\omega_{UV})^{2}}
\label{epsi_dampedAgI}
\end{equation}
Notice here that it is constructed precisely to fulfill the properties associated to any valid dielectric function: positive and decreasing function, at infinite frequencies it reaches the response of the vacuum ($\epsilon(i\xi\rightarrow\infty) =1$) and at zero frequencies it reaches the value of the static contribution ($\epsilon(i\xi\rightarrow 0) = \epsilon(0))$. Considering only the UV absorption, we can get the parameters $\omega_{UV}$ and $C_{UV}=n_{UV}^{2}-1$ from the Cauchy's representation
\begin{equation}
n^{2}-1 = (n^{2}-1)\frac{\omega^{2}}{\omega_{UV}^{2}} + C_ {UV}
\end{equation}
Using for that a linear fit with experimental data of the evolution of the refractive index at those frequencies. The other magnitudes to complete the parameterization of the Eq.   \ref{epsi_dampedAgI} were directly available in the bibliography.
\end{document}